\documentclass[acmsmall]{acmart}

\usepackage{tcolorbox}
\usepackage{lineno}
\tcbuselibrary{skins, breakable, theorems}
\newtcolorbox[auto counter]{summary}[1][]{
        title={\bfseries Summary},enhanced,
	coltitle=black,
	top=0.17in,
	attach boxed title to top left=
	{xshift=1.5em,yshift=-\tcboxedtitleheight/2},
        boxed title style={size=small,colback=lightgray},#1}

\AtBeginDocument{%
  \providecommand\BibTeX{{%
    \normalfont B\kern-0.5em{\scshape i\kern-0.25em b}\kern-0.8em\TeX}}}

\setcopyright{acmcopyright}
\copyrightyear{2023}
\acmYear{2023}
\acmDOI{10.1145/1122445.1122456}

\acmConference[Conference acronym 'XX]{Make sure to enter the correct
  conference title from your rights confirmation emai}{June 03--05,
  2018}{Woodstock, NY}
\acmPrice{15.00}
\acmISBN{978-1-4503-XXXX-X/18/06}

\acmJournal{TOSEM}
\acmVolume{X}
\acmNumber{Y}
\acmArticle{Z}
\acmMonth{2}

\usepackage{adjustbox, multicol,graphicx,mathrsfs,amsmath,pifont,amscd,latexsym,color,fancyhdr,CJK,url,longtable, pifont,epstopdf,setspace,multirow,colortbl,tabularx,threeparttable, booktabs, verbatim, bbm,listings, balance,color,indentfirst,xspace,hyperref,anyfontsize,mathtools, multirow, caption, float, subfigure}

\newcommand{\xmark}{\ding{55}}%
\newcommand{\cmark}{\ding{51}}%

\begin{document}

\title{A First Look at On-device Models in iOS Apps}

\author{Han Hu}
\email{han.hu@monash.edu}
\affiliation{%
  \institution{Monash University}
  \streetaddress{Wellington Road}
  \city{Clayton}
  \state{Victoria}
  \country{Australia}
  \postcode{3800}
}
\author{Yujin Huang}
\email{yujin.huang@monash.edu}
\affiliation{%
  \institution{Monash University}
  \streetaddress{Wellington Road}
  \city{Clayton}
  \state{Victoria}
  \country{Australia}
  \postcode{3800}
}

\author{Qiuyuan Chen}
\email{joeqychen@tencent.com}
\affiliation{%
 \institution{Tencent}
 \streetaddress{Nanshan}
 \city{Shenzhen}
 \state{Guangdong}
 \country{China}}

\author{Terry Yue zhuo}
\email{terry.zhuo@monash.edu}
\affiliation{%
  \institution{Monash University}
  \streetaddress{Wellington Road}
  \city{Clayton}
  \state{Victoria}
  \country{Australia}
  \postcode{3800}
}
  \author{Chunyang Chen}
\email{chunyang.chen@monash.edu}
\authornote{Corresponding author}
\affiliation{%
  \institution{Monash University}
  \streetaddress{Wellington Road}
  \city{Clayton}
  \state{Victoria}
  \country{Australia}
  \postcode{3800}
}

\newcommand{\chen}[1]{\textcolor{red}{#1}}

\begin{abstract}
 Powered by the rising popularity of deep learning techniques on smartphones, on-device deep learning models are being used in vital fields like finance, social media, and driving assistance.
 Because of the transparency of the Android platform and the on-device models inside, on-device models on Android smartphones have been proven to be extremely vulnerable.
 However, due to the challenge in accessing and analysing iOS app files, despite iOS being a mobile platform as popular as Android, there are no relevant works on on-device models in iOS apps.
 Since the functionalities of the same app on Android and iOS platforms are similar, the same vulnerabilities may exist on both platforms.
 In this paper, we present the first empirical study about on-device models in iOS apps, including their adoption of deep learning frameworks, structure, functionality, and potential security issues. 
 We study why current developers use different on-device models for one app between iOS and Android.
 We propose a more general attack against white-box models that does not rely on pre-trained models and a new adversarial attack approach based on our findings to target iOS's gray-box on-device models.
 Our results show the effectiveness of our approaches.
 Finally, we successfully exploit the vulnerabilities of on-device models to attack real-world iOS apps.
\end{abstract}




\maketitle

\section{Introduction}

Deep Learning (DL) models have become increasingly common in applications (apps) for smartphones in recent years~\cite{xu2019first}.
They play a vital role in the functionality features of some apps, like image classification~\cite{lu2007survey, wang2017residual, haralick1973textural}, face detection~\cite{hjelmaas2001face, rowley1998neural}, speech recognition~\cite{reddy1976speech, gaikwad2010review, povey2011kaldi}, etc.
DL models in apps are now employed in two ways: on-cloud models and on-device models~\cite{googlecloud}.
On-cloud models are stored on the remote server, triggered by instructions and data from apps via the Internet, and return results to local apps.
However, this method is heavily sensitive to the Internet's condition and may expose user privacy~\cite{el2014literature}.
Consequently, on-device models, which are put locally on the smartphone, have become a feasible option for an increasing number of apps.
With this trend, industry and academia are starting to focus on how DL models are being used in smartphone apps, including framework distribution, model attributes, potential security issues, etc~\cite{huang2021robustness, huang2022smart, madry2017towards, ali2017same,sang2023beyond}.

Researchers have systematically studied how real-world Android apps use on-device models~\cite{huang2021robustness, huang2022smart, xu2019first}.
The Android platform is an open-source ecosystem~\cite{android}.
The on-device model in \emph{.tflite} format, which is most used in Android apps, is easily accessible to third parties for the structure and the trained weights of the models~\cite{tfLite, xu2019first}.
Due to the transparency of the Android platform and the on-device models inside, on-device models on Android smartphones have been proven to be extremely vulnerable, and adversarial examples generated by common adversarial attacks can easily fool on-device models~\cite{huang2021robustness, huang2022smart}.
Attacking the models inside those apps will be disastrous for users because many mobile apps with on-device models are used for critical tasks like user authentication, medical monitoring, and driving assistance~\cite{ren2020adversarial, zhang2021deep,wang2023energy}.
Android and iOS are the two most-used platforms for smartphones~\cite{phoneSystem}.
Due to the similarity of the functionalities of the same app on the Android and iOS platforms, the same vulnerabilities may exist on both platforms~\cite{gronli2014mobile, garg2021comparative}.
However, no research has yet been done to study the state of on-device models in iOS apps.

According to our observations, there are two reasons that prevent the study of the on-device models in iOS apps.
First, Apple does not actively distribute the source files of iOS apps with third parties, and the iOS ecosystem is closed-source~\cite{appleStore}. 
The data collection and acquisition process for the iOS app is much more challenging than the Android APK files.
Second, the most used iOS-specific DL framework Core ML~\cite{coreML}, provided by Apple~\cite{apple}, is not open source either, making it challenging to study the on-device model of this framework~\cite{CoreMLConvert}.
Core ML on-device models are gray-box models that do not share trained weights with third parties and provide extra model security protections for trained weights to prevent model hacking~\cite{CoreMLConvert}.
The most effective adversarial attacks against on-device models are all white-box attacks~\cite{ren2020adversarial, kurakin2018adversarial}, which require knowing the model's structure and weights prior to the attack.
Traditional adversarial attack techniques~\cite{goodfellow2014explaining, carlini2017towards, croce2020minimally, he2019towards} perform poorly when attacking Core ML models.
These factors lead many followers to accept that iOS on-device models are by nature more 'secure' than models on Android.
Therefore, there is a lack of systematic study on characteristics and potential security flaws in on-device models on iOS.

In this paper, we present the first systematic empirical study on how real-world iOS apps exploit on-device DL models and its potential security issues.
We seek to answer three research questions: (1) what are the characteristics of on-device models on iOS; 
(2) Why developers use different models for one app on iOS and Android; 
(3) How robust are on-device models on iOS against adversarial attacks.

To answer these three questions, we first propose a pipeline for iOS app source file collection.
We collect and make public the first dataset of iOS app source files\footnote{\href{https://github.com/huhanGitHub/iOS-App-database}{{iOS-App-database}}}, which contains 2907 iOS apps from Apple App Store~\cite{appleStore} and a list of apps containing on-device models.
Following related on-device model detection approaches~\cite{xu2019first, huang2021robustness, huang2022smart}, we identify 334 iOS apps with 1883 on-device models among all the collected iOS apps.

To understand the current use of on-device models in iOS apps, we analyse the current characteristics of on-device models in 334 apps in RQ1, including their framework, model developers, model size, model functionality, and model structure.
We discover that the same app regularly adopts different DL frameworks and on-device models across Android and iOS.

To understand why developers choose different DL models for the same app on different platforms, we manually explore the iOS-Android app pairs with different on-device models in RQ2.
We summarise reasons for model changes, including platform compatibility, model usability, etc., to provide developers insights when developing mirror apps between Android and iOS.
After analysing the characteristics of the current on-device model of iOS and the similarities and differences with models on Android, we investigate the current security concerns of models on iOS.

In RQ3, we first propose a more general and effective approach to attack white-box on-device models. 
Then, based on the results of RQ1 and RQ2, we propose a new method for performing adversarial attacks against gray-box Core ML models.
We evaluate the robustness of white-box models and Core ML models against adversarial attacks in our approaches.
The experimental results reveal the effectiveness of our approaches compared with the recently proposed baseline ModelAttacker~\cite{huang2021robustness}. 
All selected 10 gray-box on-device models of Core ML are successfully attacked by our approach with an average success rate of 75\%.
Finally, we successfully use the on-device model vulnerability to fool real-world iOS apps.

In summary, our contributions are as follows:
\begin{itemize}
    \item This is the first systematic study of on-device models on iOS. 
    We collected and analysed the characteristics of the on-device models in iOS apps.
    Our research can help developers and researchers understand the current state of the on-device models in apps.
    
    \item We design and implement pipelines for analyzing on-device models on iOS and comparing on-device models between iOS and Android. Our findings show differences and motivations for selecting DL models for IOS and Android apps.
    
    \item We present a new adversarial attack approach for iOS gray-box on-device models. This is the first study on how to attack on-device models of Core ML. 
    Our method achieves an average success rate of 75\% when applying adversarial attacks on gray-box on-device models of Core ML.
    
\end{itemize}



\section{approach overview}
\begin{figure*}[htbp]
\vspace{-0.5cm}
\setlength{\abovecaptionskip}{10pt} 
\setlength{\belowcaptionskip}{10pt}
    \centering
    \includegraphics[width=\textwidth]{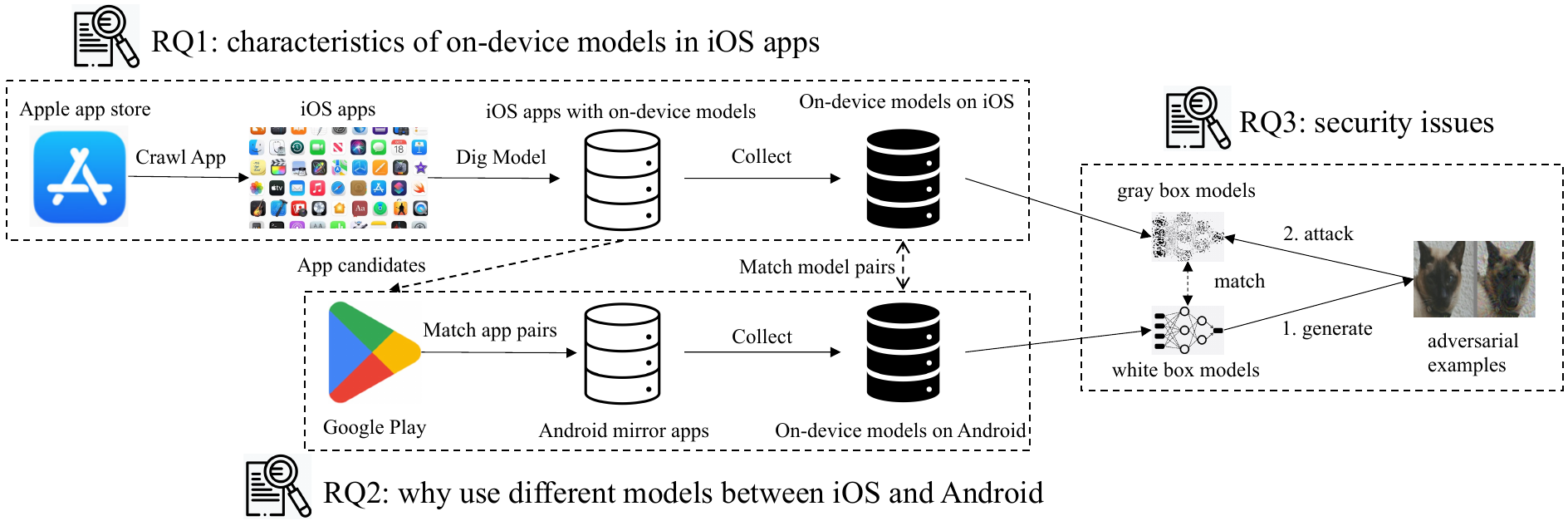}
    \caption{The workflow of this study}
    \label{fig:workflow}
    \vspace{-1cm}
\end{figure*}

We design and implement a workflow to explore our research goals.
As shown in Figure~\ref{fig:workflow}, the first step of the workflow is to crawl iOS apps from Apple App Store.
This is achieved by simulating real people downloading iOS apps on iPhone emulators using the IPA tool~\cite{ipaTool}.
Following the guideline for iOS reverse engineering~\cite{owasp}, we recompile and extract non-code resources from all crawled iOS IPA files to detect on-device models.
We study these apps and on-device models in RQ1.
Second, we match identified iOS apps with their Android mirror apps on Google Play~\cite{googleplay}.
We study how and why developers choose different on-device models across Android and iOS.
Third, we propose a new approach for employing adversarial attacks to gray-box Core ML models on iOS.
We employ our proposed adversarial attack approach to evaluate the robustness of gray-box on-device models on iOS in RQ3.
Finally, we select attacked models to identify the point in apps where models are invoked and then manually input the adversarial examples to validate our method's effectiveness in directly attacking real-world iOS apps.

We will discuss the detail of crawling apps, detecting on-device models, paring Android counterparts, and attacking models in the following sections.

\section{RQ1: what are the characteristics of on-device models in iOS apps}

\subsection{Motivation}
As with the study of on-device models in Android~\cite{huang2021robustness, huang2022smart, xu2019first}, this RQ focuses on on-device models' existing characteristics in iOS apps.
To comprehend the current trends, characteristics, and experiences of on-device models employed by iOS developers, we concentrate on DL framework selection, model quantity, model size, model type, model functionality and model developer for iOS app models.
The exploration of this RQ can support our future investigation into the causes of the current trends and the potential security vulnerabilities associated with the current trends.

This is the first study of the on-device model in iOS, to offer potential insights to future researchers, we begin by presenting our pipeline for data collection in the iOS platform, including how to acquire iOS app files from the Apple App Store~\cite{appleStore} and how to recognise on-device models in iOS apps.

\subsection{iOS App Collection}
\label{sec:appCollection}

\subsubsection{How to Select Apps with On-device Models}
First, we download all 2312 top-rated free apps across 25 categories which are all publicly available on the Apple App Store~\cite{topRatedApp}.

Second, we review the literature on common application scenarios of DL models in mobile apps~\cite{gcloud, xu2019first} and identify 16 prevalent applications based on our own expertise and observations. 
These applications include photo enhancement, object detection, image classification, face detection, image segmentation, optical character recognition (OCR) text recognition, augmented reality, barcode scanning, language identification, smart replies, translation, sound recognition, recommendations, movement tracking, video segmentation, and gesture recognition.
To identify relevant apps, we conduct a search on the iOS app store.
We utilize the identified application scenarios as search keywords and manually sift through the search results to isolate apps that exhibit these features. 
Our search yields 205 apps that align with the aforementioned application scenarios in this way.

Third, one app is usually available on both Android and iOS platforms~\cite{ali2017same}.
Due to their similar functionality, iOS apps may contain on-device models if on-device models are discovered in their Android counterparts.
We discover 423 Android apps that contain on-device models out of a total of 26,346 Android apps crawled from Google Play~\cite{googleplay}.
To match their iOS counterparts, we start by finding corresponding iOS apps with the same app names and developers.
However, the name of the same app may be slightly different on different platforms~\cite{ali2017same}.
For instance, the \emph{Tubi} app's official name on Android is \emph{Tubi-Movies \& TV Shows} whereas it is \emph{Tubi-Watch Movies \& TV Shows} on iOS.
Therefore, we manually locate iOS equivalents for Android apps that cannot be found with the same app name and developer.
In this way, we collects 390 iOS apps.

Until September 30, 2022, we collected a total of 2,907 iOS apps by using previous three approaches.

\subsubsection{How to Get Source Files of iOS Apps}
After logging in with an apple id, we use automated python scripts to imitate the user downloading all free apps from the apple store to the iPhone emulator. Then, we utilise the IPA~\cite{ipaTool} tool to extract all of the downloaded apps' source files.
We make all our collected iOS apps publicly available for more researchers to study\footnote{\href{https://github.com/huhanGitHub/iOS-App-database}{iOS-App-database}}.

\subsection{On-device Model Collection}

\begin{table*}[htbp]
\vspace{-0.5cm}
\setlength{\abovecaptionskip}{10pt} 
\setlength{\belowcaptionskip}{10pt}
\caption{An overview of popular deep learning frameworks and their smartphone support at the time of writing (Sep. 2022).}
\begin{adjustbox}{ width=\textwidth,center}
\centering
\begin{tabular}{|lcccccc|}
    \noalign{\hrule height 1pt}
\textbf{Framework} & \textbf{Owner} & \textbf{Mobile Platform} &
\textbf{Mobile API} & \textbf{Is Open-source} & \textbf{Supported Model Format} & \textbf{Support Training} \\ 

    \noalign{\hrule height 1pt}

ONNX~\cite{ONNX} & ONNX & Android \& iOS & Java, C, C++, OC & \cmark & Protobuf (\emph{.pb, .onnx}), Numpy (\emph{.npz}) & \cmark\\ 

TF Lite~\cite{tfLite} & Google & Android \& iOS & Java, Python, C++, OC, Swift & \cmark & FlatBuffers (\emph{.tflite}) & \cmark\\ 

Caffe~\cite{caffe} & Berkeley & Android \& iOS & C++ & \cmark & customized, json (\emph{.caffemodel, .prototxt}), json, YAML & \cmark\\ 

Caffe2~\cite{caffe2} & Facebook & Android \& iOS & C, C++ & \cmark & ProtoBuf (\emph{.pb}) & \cmark\\ 

MxNet~\cite{mxnet} & Apache Incubator & Android \&  iOS & C, C++ & \cmark & customized, json (\emph{.json, .params}) & \cmark\\ 

DeepLearning4J~\cite{dl4j}  & Skymind & Android \& iOS & Java & \cmark & customized (\emph{.zip}) & \cmark\\ 

ncnn~\cite{ncnn} & Tencent & Android \& iOS & C++ & \cmark & customized (\emph{.params, .bin}) & \xmark\\ 

OpenCV~\cite{opencv} & OpenCV Team & Android \& iOS & C, C++ & \cmark & TensorFlow, Caffe, Darknet, ONNX, Torch, PyTorch & \cmark\\ 

FeatherCNN~\cite{feathercnn} & Tencent & Android \& iOS & C++ & \cmark & customized (\emph{.feathermodel}) & \xmark\\ 

Paddle-Lite~\cite{pdlite} & Baidu & Android \& iOS \& Kirin  & Java, C++ & \cmark & parambase (\emph{.pdmodel, .pdparams, .pdopt}), TensorFlow, Caffe, ONNX, PyTorch & \xmark\\ 

MNN~\cite{mnn} & Alibaba & Android \& iOS & Python, C++ & \cmark & TensorFlow, Caffe, Darknet, ONNX & \cmark\\ 

MACE~\cite{mace} & XiaoMi & ARM-based \& Android \& iOS & C++ & \cmark & customized (\emph{.pb, .yml, .a}), TensorFlow, Caffe,  ONNX & \xmark\\

CoreML~\cite{coreML} & Apple & iOS \& iPadOS \& watchOS & OC,Swift & \xmark & customized, ProtoBuf (\emph{.proto, .mlmodel}), TensorFlow, Caffe,  ONNX, PyTorch & \cmark\\ 

PyTorch Mobile~\cite{pytorch-mobile} & Facebook & Android \& iOS & Java, OC, Swift & \cmark & customized, pickle (\emph{.ckpt, .pkl, .pt, .pth, .ptl}), ONNX & \xmark\\ 

Bender~\cite{bender} & Xmartlabs & iOS & Swift & \cmark & Tensorflow & \xmark\\ 

    \noalign{\hrule height 1pt}
\end{tabular}
\end{adjustbox}
\label{tab:overview_of_frameworks}
\vspace{-0.5cm}
\end{table*}

Inspired by related on-device model detection approaches~\cite{xu2019first, huang2021robustness, huang2022smart}, we identify the on-device model by matching specific suffix patterns of the model files in reverse-engineered iOS app source files.

We first investigate popular DL frameworks today that enable the deployment of iOS on-device models.
Following the framework selection process in related works~\cite{xu2019first, huang2021robustness}, we commence our investigation by searching Google and GitHub to identify popular DL frameworks.
We collect relevant information on the features of these frameworks from their official documentation, such as the supported model formats, the suffix patterns of the on-device model supported, open-source status, etc.
We also adopt a similar set of evaluation criteria employed in prior studies~\cite{xu2019first, huang2021robustness} to determine the suitability of the identified frameworks. 
Specifically, frameworks that have not been actively maintained for more than two years and have garnered minimal attention, such as fewer than 100 open-source projects on GitHub, are excluded from consideration in this study.
Table~\ref{tab:overview_of_frameworks} shows the overview of this investigation.
We delete frameworks with fewer than 100 GitHub stars and no updates in the past two years~\cite{xu2019first}.
As illustrated in Table~\ref{tab:overview_of_frameworks}, the columns \emph{Framework}, \emph{Owner}, \emph{Mobile Platform}, \emph{Mobile API} and \emph{Supported Model Format} represent the names, owners, supported mobile platform, supported mobile API and supported model format of DL frameworks, respectively.  
We refer to the suffix patterns in the column \emph{Supported Model Format} to detect on-device models.
The columns \emph{Is Open-source} and \emph{Supported Training} indicate whether the framework is open source and whether the framework's on-device model supports training. 
There are 15 popular DL frameworks support deploying on-device models on iOS now.
9 out of 15 framework models support training, while the remaining models only support prediction using trained models.

To ensure the quality of the model and remove false positive cases, we evaluate the quality of each detected model by loading the model and performing predictions on randomly generated data.
Finally, we discover 1,883 valid on-device models among 2,907 iOS apps after eliminating all non-predictive models.
These 1,883 on-device models belong to 334 iOS apps, of which 190 apps are from top-rated free apps, 40 are from related application scenarios searches, and 104 are from matching Android-iOS app pairs.

\subsection{Characteristics of Apps with On-device Models}


We study the characteristics of existing on-device models on the iOS platform from two aspects. 
The first aspect is to look at the features of apps containing on-device models, including the category of apps in Section~\ref{sec:appCate}, the number of models contained in each app in Section~\ref{sec:appNum}, the size of the model as a percentage of the app's size in Section~\ref{sec:modelSize} and the developers of DL apps in Section~\ref{sec:appDev}.
Apps dominant portion of smartphones and platforms~\cite{android, apple}.
The first perspective enables us to understand current app-level trends in the employment of on-device models on the iOS platform.
The second aspect is to look at the features of on-device model itself, including the type of current models in Section~\ref{sec:modelType}, the functionality of current models in Section~\ref{sec:modelFunction} and the adoption of DL frameworks in Section~\ref{sec:frameCompare}.
The second perspective allows us to gain insight into the model-level characteristics, laying the foundation for the subsequent research questions.

\subsubsection{How On-device Models are Distributed among App Categories?}
\label{sec:appCate}
We explore the categories of  334 DL model apps.
The top five categories of apps are Photo \& Video (44), Shopping (33), Social Networking (22), Health \& Fitness (20), and Travel (17).
Note that most on-device models are still used in computer vision-related scenarios, like OCR text detection, object recognition, and face detection, even though these apps fall outside of Photo \& Video category.
The findings show that computer vision-related scenarios still predominate the industrial applications of deep learning.

\subsubsection{How Many On-device Models are there in One App?}
\label{sec:appNum}
\begin{figure}[htbp]
\vspace{-0.5cm}
\setlength{\abovecaptionskip}{10pt} 
\setlength{\belowcaptionskip}{10pt}
	\centering
	\subfigure[The number of models in apps]{
		\begin{minipage}[t]{0.45\linewidth}
		\setlength{\abovecaptionskip}{0pt} 
\setlength{\belowcaptionskip}{10pt}
			\centering
		\includegraphics[width=\linewidth]{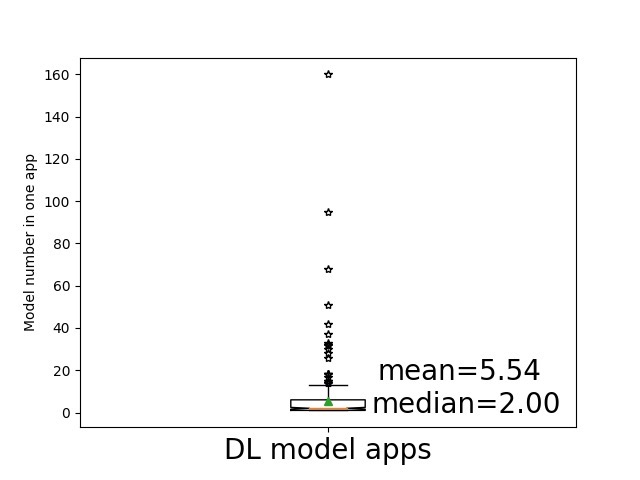}\\
			\label{fig:modelNum}
		\end{minipage}%
	}%
	\subfigure[The proportion of model sizes in apps]{
		\begin{minipage}[t]{0.45\linewidth}
		\setlength{\abovecaptionskip}{0pt} 
\setlength{\belowcaptionskip}{10pt}
			\centering			\includegraphics[width=\linewidth]{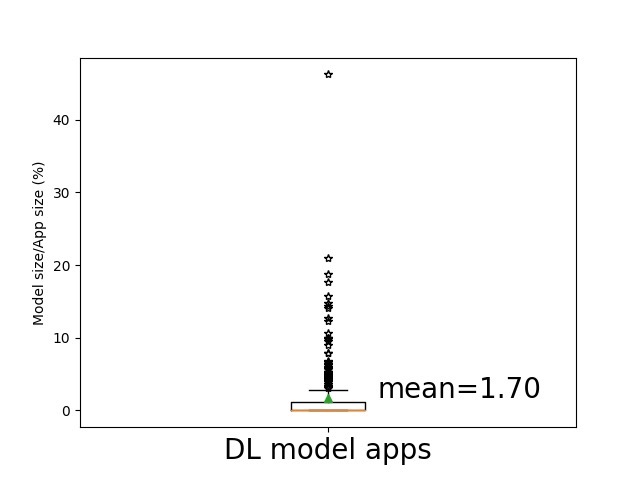}\\
			\label{fig:modelSize}
		\end{minipage}%
	}%
	\caption{The boxplot of distributions of the number of on-device models and the proportion of model sizes in iOS apps}
	\label{fig:modelComparision}
	\vspace{-0.5cm}
\end{figure}

Figure~\ref{fig:modelNum} shows the boxplot of the number of on-device models in one app on iOS.
The mean is 5.54, and the median is 2, indicating that the average number of models in a single app is 5.54 and that half of DL model apps have less than 2 models.
The outliers in the box plot demonstrate that 14 apps own more than 20 DL models.
These outliers are all photo/video processing apps, like Gradient (95), Video Star (51), Facetune2 Editor (160), etc.
These apps provide a multitude of image-processing features, such as skin smoothing, eye enlargement, beauty, etc.
To fulfill these features, developers typically employ multiple pre-trained on-device models.
Hence, these apps contain more on-device models.

\subsubsection{What is the Size of On-device Models?}
\label{sec:modelSize}
Given the limited computing and storage resources of mobile phones, the size of the model is a crucial indicator of its suitability for deployment on such devices. Models with large sizes may cause performance and storage issues, while smaller models may be more practical for use on mobile devices~\cite{onCloud, xu2019first}.
The average model size of detected 1,883 on-device models is 0.45MB.
Figure~\ref{fig:modelSize} demonstrates the boxplot of the percentages of the app's size that is occupied by the app's models (Total model sizes/App size).
Results show that the present app's model size is barely average of 1.7\% of the app size.
The apps with the highest proportion are \emph{SoundLab Audio Editor} 46.26\%, \emph{iScanner} 20.92\% and \emph{Carl: Plant Identification} 18.77\%.
The average proportion of on-device models is small, but outliers in Figure~\ref{fig:modelSize} illustrates that the proportion of model size exceeds 10\% in all apps where deep learning plays a crucial role, such as \emph{B612}'s 17.63\%, \emph{EPIK Photo Editor}'s 14.37\%, and \emph{VITA Video Editor}'s 15.72\%.
Even though the average proportion of model size is quite small, the proportions of models in machine learning-related apps are considerable.
Large models in apps will increase the cost of development and maintenance and may have a negative effect on the user experience if they slow down the app~\cite{ballard2007designing}.
Optimizing and downsizing the on-device models is an urgent and promising direction.

\subsubsection{Who Develops On-device Model Apps?}
\label{sec:appDev}
The iOS app store provides a details web page for each app, which includes relevant information about the app's developer. 
In this study, we crawl these web pages and collect the developer information for 334 apps.
334 iOS apps are developed by 297 different companies or developers. 
\emph{Google LLC} has the most apps that employ on-device models, with 10, followed by \emph{Microsoft Corporation} with 5 and \emph{Meta Platforms} with 4.
We notice that Google's apps, such as \emph{Google Street View}, \emph{Google Family Link}, and \emph{Google Home}, typically employ their own frameworks: TensorFlow and TF Lite.
Correspondingly, Microsoft, Meta, and the majority of small- and medium-sized companies, such as Zoom, Netflix, and SHEIN, commonly employ the third-party frameworks Core ML and TF Lite in their apps, such as \emph{Microsoft Office}, \emph{Microsoft OneDrive}, \emph{Meta Business Suite}, and \emph{Oculus}.
Owing to the accumulation of technology and vast data,
large tech companies are the most prolific developers of on-device models.

\subsubsection{What Types are Current On-device Models?}
\label{sec:modelType}
Among 1,883 on-device models, 1,561 (82.9\%) models are CNN models, which are primarily used for image and video classification and detection, 163 (8.7\%) models are RNN models, which are primarily used for text and sound classification, and remaining 159 (8.4\%) models fail to identify the model structures.
The results are consistent with the model functionality analysis in Section~\ref{sec:modelFunction}, as the majority of on-device models in current iOS apps are used in the computer vision domain, and hence the majority of models are of type CNN.

\subsubsection{What are On-device Models Used for in iOS Apps?}
\label{sec:modelFunction}
To figure out what on-device models are used for in iOS apps, we propose a strict pipeline to analyze the functionalities of current on-device models.

\begin{figure*}[htbp]
	\vspace{-0.5cm}
\setlength{\abovecaptionskip}{10pt} 
\setlength{\belowcaptionskip}{10pt}
	\centering
    \begin{minipage}[h]{0.5\linewidth}
    \centering
    \captionof{table}{The usage of on-device models in iOS apps}
    \begin{tabular}{|c|c|c|}
        \hline
        \textbf{Field} & \textbf{Usage} & \textbf{Count} \\
        \hline
        \multirow{8}*{CV: 332}  & Photo beauty & 68 \\ 
        
        ~ & Object detection & 48 \\ 
        ~ & Image classification & 46 \\ 
        ~ & Face detection & 39 \\ 
        ~ & Image segmentation & 37 \\ 
        ~ & OCR Text recognition & 36 \\
        ~ & Augmented reality & 36 \\ 
        ~ &  Barcode scanning & 22 \\ 
        
        \hline
        
        \multirow{3}*{NLP: 23} & Language identification & 20 \\ 
        
        ~ & Smart replies & 2 \\
        
        ~ & Translation & 1 \\ \hline
        
        Audio: 10 & Sound recognition & 10 \\ \hline
        
        \multirow{5}*{Other: 55} & Recommendation & 17 \\
        
        ~ & Unknown & 15 \\
        
        ~ & Movement tracking & 13 \\
        
        ~ & Video segment & 5 \\
        
        ~ & Gesture recognition & 5 \\
        \hline
    \end{tabular}
    \label{tab:modelUsage}
	\end{minipage}%
	\begin{minipage}[h]{0.48\linewidth}
    \centering
    \includegraphics[width=\linewidth]{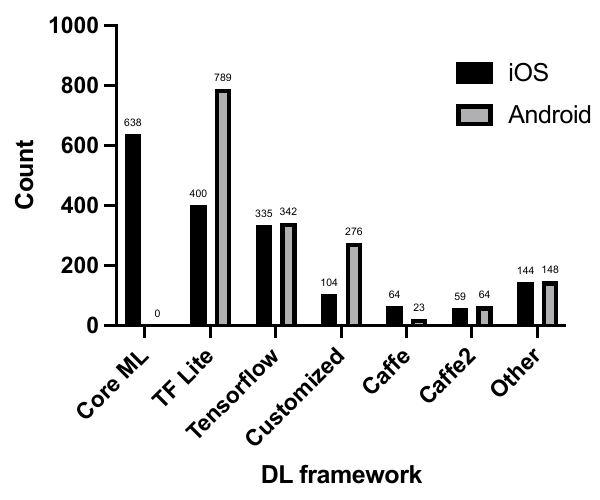}
    \vspace{-4mm}
    \caption{Comparison of the distribution of DL frameworks on iOS and Android }
    \label{fig:dlDis}
	\end{minipage}%
	\vspace{-0.2cm}
\end{figure*}

We classify the functionality of a model in three steps.
According to a recent study~\cite{huang2021robustness}, over 60\% of existing DL models in industrial apps are structurally comparable to existing pre-trained models from TensorFlow Hub~\cite{tfhub}.
So, we first manually check whether the model is fine-tuned from typical model architectures, such as ResNet~\cite{he2016deep}, Mobilenet~\cite{howard2017mobilenets}, YOLO~\cite{redmon2016you}, etc.
Typical model architectures are used for specific tasks, such as YOLO for detection, ResNet for classification, and DeepLab~\cite{chen2017deeplab} for segmentation.
The model's functionality could be identified by its classical model structure.
Following the related work proposed by Huang et al.~\cite{huang2021robustness},  we assess whether a given model has been fine-tuned from one of the models in TensorFlow Hub by comparing the structure and parameter similarity between the two models. 
Specifically, we first extract the structure information of the model in terms of layer names, shapes, and data types, and represent it as a string sequence. 
Similarly, we convert a set of pre-trained models into string sequences. 
We then compute the similarity between the model sequence and the pre-trained model sequences. 
Based on this similarity, we can determine whether the model's structure is similar to that of any pre-trained models. 
We follow related works~\cite{huang2022smart, sim2019investigation} to set the threshold of similarity to 80\%.
If the model structure matches a pre-trained model, we compute the parameter similarity between the two models to verify whether the model has been fine-tuned.
Second, we infer the use of models by analyzing the output layer of the mode.
Models with a single activation function, such as \emph{Softmax} and \emph{Sigmoid}, are typically employed for classification tasks\cite{activationFunction}.
Due to the necessity of locating the object's bounding box, the detection task's output layer of models contains regression operations~\cite{szegedy2013deep}.
After the previous two steps, we can identify the model's functionality at a high level.
In the final third step, we follow the approach proposed by Xu et al.~\cite{xu2019first} to infer the model's precise usage scenario by analyzing the semantic data in the app, including the app description, app content, the model name, the layer name, detected labels, and other app-related information.
Finally, we run the app to validate our inferences.
For instance, if the model name is \emph{face\_detect.tflite}, we first confirm that it is a detection model by analyzing the last layer of the model. Next, we infer from the model name and the app's documentation that the model may have a face detection function. Since face detection requires the use of the camera and pictures, we then search for all places in the app that use the camera and pictures for content recognition to verify if face detection exists. If we find face detection in any of these places, we assume that the model's function is face detection. However, if the face detection function is not found, we assign the model's function to object detection at a high level.

Given the amount of effort, we randomly sample 68 (20\%) apps in the 334 iOS apps to study their model functionalities.
We discover 420 (22\%) on-device models among these 68 apps.
Table~\ref{tab:modelUsage} shows the results of our model functionality analysis among these 420 models.
332 (79\%) on-device models are used in the computer vision field.
The most widely used scenario is Photo Beauty (68), followed by Object Detection (48) and Image Classification (46).
In the natural language processing field, we discover that Language identification (20) is the scenario that uses on-device models the most.
On-device models like \emph{tflite\_langid.tflite} are frequently used by apps like \emph{Camera Translator: Translate +}, \emph{Think Dirty - Shop Clean}, and \emph{Wizz - Make new friends} to distinguish the types of input languages.
Sound recognition (10) is the most widely-used task in the audio field.
In addition, on-device models are also frequently utilized for recommendations (17), particularly in planning apps like \emph{Structured-Daily Planner}, \emph{Motivation-Daily quotes}, and \emph{I am-Daily Affirmations}, where developers frequently make use of the trained On-device model to provide users with appropriate recommendations.

According to the results, computer vision-related domains continue to be the most popular application areas for on-device models. Despite the widespread usage of deep learning techniques in machine translation, developers currently prefer to employ on-cloud deep learning models or third-party APIs for translation and on-device models to identify the type of input language.

\subsubsection{What is the Adoption of DL Frameworks in iOS Apps? How does it Compare to Models in Android Apps?}
\label{sec:frameCompare}

We investigate the adoption of DL frameworks in 334 iOS apps with on-device models and compare it with their Android counterparts.
1,744 and 1,642 on-device models are validated successfully for the frameworks on 334 iOS and Android apps, respectively.

Figure~\ref{fig:dlDis} demonstrates the distribution of DL frameworks in 334 iOS apps and their comparisons on Android.
There is a significant difference between the distribution of DL frameworks in Android and iOS.
The most widely-used framework in iOS is Core ML, with 638 (36.58\%), followed by TF Lite, with 400 (22.94\%), and TensorFlow 335 (19.21\%).
The first three frameworks combined make up over 75\% of the market for industrial apps.
Android apps usually employ TensorFlow and TF Lite as DL frameworks, accounting for 789 (48.05\%) and 342 (20.83\%) out of 1,642 DL models.
Compared to iOS apps, Android apps have more customized DL models (276, 16.81\%). 
One possible explanation is that Core ML and TF Lite are optimized for their respective platforms and exhibit better compatibility, resulting in lower development costs. Additionally, both frameworks have a large community of users and developers who provide support and resources, making them more accessible and easier to use.
Moreover, the popularity of these frameworks may also be attributed to the companies behind them. Apple and Google are well-established and widely recognized tech giants, and their frameworks may be perceived as more reliable and trustworthy compared to other third-party alternatives.

Moreover, we find that TF Lite and TensorFlow remain the second and third most used frameworks in the iOS platform, with a share of 22.94\% and 19.21\%, respectively. This suggests that the continued popularity of TF Lite and TensorFlow in the iOS platform is not solely based on their ownership by Google but may also be due to other factors.
One possible reason for their popularity is that TensorFlow has been in the market longer than Core ML and has established itself in the deep learning community. Developers may have more experience and expertise in using TensorFlow, making it easier for them to integrate it into their iOS apps. Furthermore, TensorFlow has a larger user base, providing developers with a larger community for support and resources.
Another reason for the popularity of TensorFlow is its wide range of functionalities, including support for natural language processing and speech recognition, in addition to computer vision tasks. This versatility means that TensorFlow may be preferred by developers who require support for a wider range of tasks. Furthermore, TensorFlow may have cross-platform compatibility, allowing developers to use the same model across multiple platforms, including iOS and Android.
We discover that these customized models are primarily derived from three sources: customized DL SDK, like SenseTime~\cite{sstimeSDK}, traditional machine learning frameworks like XGBoost~\cite{xgboost} and compressed/obfuscated DL models.
Given that Android is considered to be less safe than iOS, more Android developers choose to compress or encrypt the model's parameters and structure to improve the security of on-device models~\cite{huang2021robustness}.
Compared to Android apps, iOS apps are more likely to utilize Caffe.
However, possibly due to the development expense, many iOS apps, which employ the Caffe framework, such as \emph{NETGEAR Nighthawk - WiFi App} and \emph{Stash: Invest \& Build Wealth}, use the more popular frameworks TF Lite and TensorFlow in their corresponding Android apps, rather than Caffe.
A tiny number of Android and iOS apps use other frameworks like NCNN~\cite{ncnn}, Paddle Lite~\cite{pdlite}, etc.
Mainstream apps tend to employ popular and stable DL frameworks like Core ML, TF Lite, and TensorFlow.

On the developer side, reusing the same framework technology when migrating apps across platforms can reduce development costs and improve development efficiency.
On the model side, popular deep learning frameworks support both Android and IOS platforms, and on-device models are intentionally designed to be highly re-usable~\cite{coreML, TensorFlow, xgboost, caffe2, mxnet}.
Despite the framework's efforts for cross-platform reuse architecture, current developers have not adopted it to its fullest extent.

\begin{summary}{}{}
On-device models are utilized extensively across all categories of apps for various functions, especially photo beauty and object detection.
Framework vendors should consider optimizing the model size and quantity in apps.
Large tech companies prefer to on-device models in their apps.
Current on-device models are mostly used in the field of computer vision, hence the majority of models are of the CNN variety.
Developers are more likely to utilize DL frameworks provided by Android or iOS operating system owners like Core ML, TF Lite and TensorFlow.
The consistency between the same app on the Android and iOS platforms utilising the DL Framework is glaringly lacking.
\end{summary}

\section{RQ2: Why developers use different models for one app on iOS and Android?}
To conquer the market, one mobile app can be available on both Android and iOS~\cite{ali2017same}.
As discussed in Section~\ref{sec:frameCompare}, considering the cost of development associated with retraining a model, most developers should accept reusing models when developing mirror apps for other platforms.
However, we discovered notable differences of DL framework selection between Android and iOS apps in Section~\ref{sec:frameCompare}.
To investigate the impact of platform differences on the selection of DL frameworks by developers, we analyze the sharing and replacement of on-device models in iOS-Android app pairs and conclude why this occurs.

In this study, we first provide the method for comparing DL models to locate the same DL models between two platforms in Section~\ref{sec:modelCompare}.
We investigate current models' cross-platform selection in Section~\ref{sec:modelChange}.
Then we design a pipeline to investigate how current developers replace and share on-device models and analyze the underlying reasons for adopting different DL models in Section~\ref{sec:factorStudy}.
Through the study of RQ2, we explain how and why current developers use different DL models to support other developers in developing Android and iOS mirror apps.
We also provide suggestions for subsequent cross-platform optimization directions for the current DL frameworks from a developer's perspective.
It establish the groundwork for RQ3 to further analyse if current cross-platform usage approaches of on-device models pose any security threats.

\subsection{Model Comparison}
\label{sec:modelCompare}
We first match the same models between the two platforms.
Model structure and trained model parameters are the primary distinguishing characteristics of DL models~\cite{goodfellow2016deep}.
So, following related works~\cite{huang2021robustness, huang2022smart}, we identify reused on-device models by comparing their structure similarity (SS) and trained parameter similarity (PS).
Two model pairs are considered the same only if the structure and parameters are identical, as denoted by SS and PS values of 1.

\subsubsection{How to Compare Models' Structure Similarity?}

We convert an on-device model to a sequence of elements, with each element constructed by the shape, data type, and name of each layer from the model~\cite{huang2021robustness, huang2022smart}. 
Given model $M_1$ and model $M_2$, their structural similarity is the longest common subsequence between the two models.
It is calculated by
\begin{equation}
    SS(M_1, \ M_2) = \frac{2 * L_{l}}{L_{M_1} + L_{M_2}}
\end{equation}
where $L_{l}$ is the length of the longest common subsequence in two models, and $L_{M_1}$ and $L_{M_2}$ are the numbers of two models' layers. 
Note that two layer elements are matched only if their attributes are totally the same.
The similarity score ranges from 0 to 1, and the higher the value, the greater the structural similarity between the two models.

\subsubsection{How to Compare Models' Parameter Similarity?}
\label{sec:ps}
After comparing the model structure similarity, we extract the trained parameters in each layer to compare the trained parameter similarity.
We count the number of elements with the same parameter in the longest common subsequence as their parameter similarity~\cite{goodfellow2016deep, huang2021robustness, huang2022smart}:
\begin{equation}
    PS(M_1, \ M_2) = \frac{2 * N_{l}}{L_{M_1} + L_{M_2}}
\end{equation}
where $N_{l}$ is the number of the same parameter elements in their longest common subsequence.

\subsubsection{How to Compare Gray-box Models?}
\label{sec:ps2}
As shown in Figure~\ref{fig:dlDis}, the gray-box model of Core ML framework is the most frequent model in iOS apps, accounting for more than one-third of the total.
For better platform compatibility, Apple encourage developers to convert the trained on-device model into the format of Core ML framework~\cite{CoreMLConvert}, such as converting the format from TF Lite in Android to Core ML in iOS.
Due to the same structure and parameters, the converted on-device and original models are considered equal.
However, it is hard for third parties to extract the trained parameters of Core ML models~\cite{coreML}.
To compare the Core ML model with the white-box model, we perform the following procedures.

We attempt to convert all eligible models to the Core ML format and compare the structure, weight and metadata of the converted models to verify if they are consistent.
Faced with models which cannot be converted to Core ML format, we first compare the structural similarity of the models and select two models with the same structure for further comparison.
We prepare a set of input data to feed to both models for prediction.
Two models are considered the same if their outputs are exactly the same.

\subsection{Model Selection on Two Platforms}
\label{sec:modelChange}
To acquire a comprehensive understanding of the present differences in DL model selection between iOS and Android developers, we first analyse quantitatively, among 334 matched iOS-Android app pairs, how many apps adopt distinct on-device models on Android and iOS.
We follow steps in Section~\ref{sec:modelCompare} to automatically and manually compare all models in 334 iOS-Android app pairs.

\subsubsection{How Many Apps Adopt Different Models?}
In 334 corresponding Android apps, 125 (37.43\%) use completely distinct on-device models from that comparable iOS apps.
68 (20.36\%) out of 334 Android apps do not use on-device models.
Only 22 (6.59\%) out of 334 Android apps have on-device models that are all the same as their iOS counterparts.
21 (6.29\%) Android apps share models in part with their corresponding iOS apps.
These iOS-Android app pairs reuse a portion of on-device models across both platforms.
Our finding shows that most apps adopt different on-device models and frameworks on different platform apps.

\subsubsection{How Many Models are Shared between iOS and Android?}
\label{sec:ratio}
We match 332 (17.63\% ) the same on-device model pairs out of 1,883 on-device models on iOS.
We find 270 (81.33\%) out of 332 model pairs are exactly the same, including structure, trained parameters, and DL framework.
62 (18.67\%) out of 332 model pairs have the same structure and parameters but different DL frameworks.
On Android, these 62 on-device models are in TF Lite framework format, whereas on iOS, it is converted to gray-box Core ML model.


\subsection{Reasons for Model Selection}
\label{sec:factorStudy}

\begin{figure*}[htbp]
\vspace{-0.5cm}
\setlength{\abovecaptionskip}{10pt} 
\setlength{\belowcaptionskip}{10pt}
    \centering
    \includegraphics[width=0.9\textwidth]{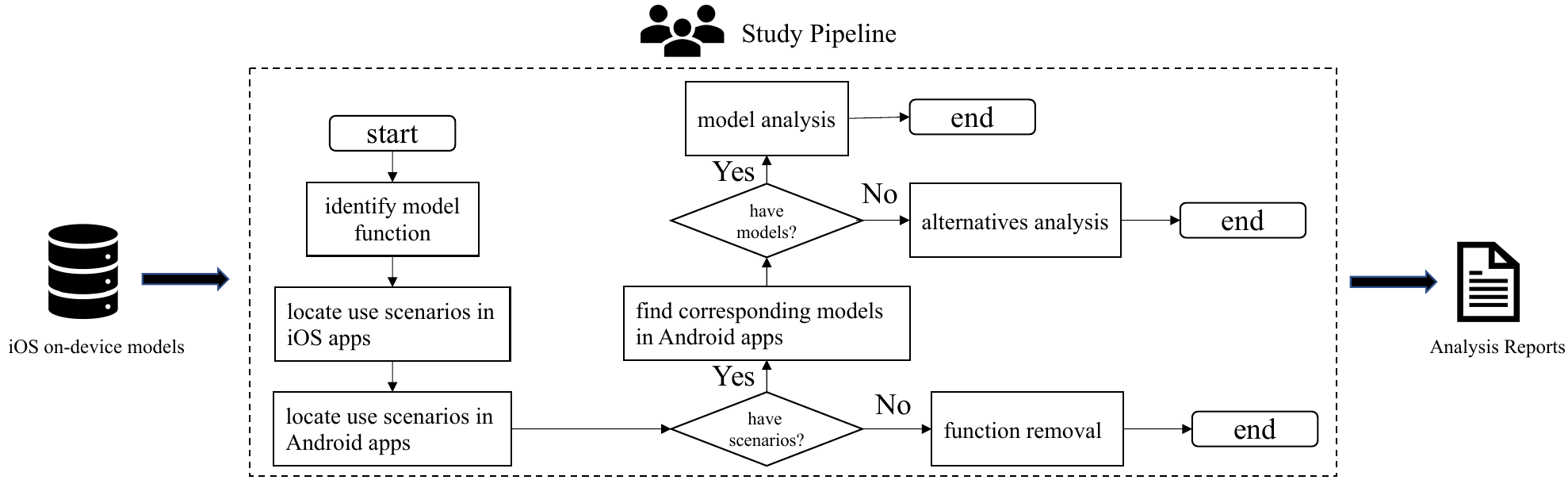}
    \caption{The study pipeline of exploring reasons for model selection}
    \label{fig:rq2}
    \vspace{-0.5cm}
\end{figure*}

We provide a strict pipeline to study how current developers replace and share on-device models between Android and iOS and the underlying motives.
Considering the tremendous workload, we manually study a statistically representative random sample of 62 (20\%) out of 312 iOS-Android app pairs for a detailed study.
We find 226 on-device models with identified scenarios and functionalities in 62 iOS apps.

\subsubsection{Study Pipeline}
Figure~\ref{fig:rq2} shows the study pipeline of the effecting factors.
First, we identify the functionalities and usage scenarios of on-device models in selected iOS apps.
Second, we attempt to locate these use scenarios and functionalities within Android counterparts.
In this step, we follow the methodologies employed by recent related works~\cite{xu2019first, chen2021my, huang2022smart} to analyze the source code and semantic information in our investigated mobile apps.
For Android apps that can accurately find the code that calls the model, we instrumented the method at which the model is called and then dynamically run the app to dynamically analyze the app usage scenario where the model is called.
In cases where the calling code of the model could not be located, we manually analyze the semantic information of the model's functionality, file names, code comments, method names, and the user interface in the app to infer the possible scenarios in which the model could be used in the app. Finally, we validate our inferences by running the app and verifying whether the inferred scenarios encompassed the model's functionalities.
If we cannot find comparable functionality or usage scenarios in Android apps, we conclude that the original functionality of the on-device model on iOS has been discarded in their Android counterparts.
If we find the corresponding functionality and model, we further compare whether the model is the same.
If comparable functionality and usage scenarios can be found, but no corresponding on-device model, the role of the model on the iOS app is regarded as having been replaced in Android counterparts.

There are now three alternatives for on-device models: using other on-device models for the same task, invoking on-cloud models~\cite{gcloud, onCloud}, and implementing custom techniques to obfuscate models.
We first check if there are other models in the app that provide the required app functionality.
According to guidelines of the current frameworks for invoking on-cloud models~\cite{onCloud}, developers need pre-configure the cloud-related SDK libraries in the configuration files of the Android apps before invoking on-cloud models in the code.
For example, developers must pre-configure \emph{play-services-mlkit-text-recognition} in the configure file to invoke on-cloud text recognition services.
So, we first check if there are configured cloud-related SDK libraries in the configuration files of the Android app.
Then, we statically analyze whether apps invoke on-cloud models via application programming interfaces (APIs) provided by DL frameworks~\cite{mlKit} in decompiled app source files.
After failing to find the relevant on-device or on-cloud models, we inspect the decompiled Android apps to manually identify alternatives and customized DL models.
We search for two file types: the lib file of the DL frameworks associated with the model's functionality and the processed DL model file. 
Apps that employ deep learning techniques should include related libs~\cite{xu2019first, mlKit, pytorch-mobile}.
For example, the library file \emph{libtensorflowlite\_jni.so} is included in the app which uses TensorFlow.
Therefore, we used the presence of these files as an indicator of the usage of DL frameworks in the app.
We filter the obfuscated model file by analyzing the semantic information associated with the model, such as the names of the obfuscated and source models, and the names of the code files that call the model. 
If both the library file and the obfuscated model file are identified, we infer that the app's developer used a customized technique to obfuscate the model.

We invited three industry developers with at least one year of experience in Android and iOS development to conduct this study.
Each of the three participants follow the above steps to inspect all selected iOS-Android app pairs to explore the causes of change.
After the initial inspection, three volunteers have a discussion and merge conflicts.
They clarify their findings, scope boundaries among categories, and misunderstandings in this step.
Finally, they iterate to revise study results and discuss with each other until a consensus is reached.

\begin{table*}[htbp]
\vspace{-0.5cm}
\setlength{\abovecaptionskip}{10pt} 
\setlength{\belowcaptionskip}{10pt}
    \caption{Results of model sharing and replacement in iOS-Android app pairs}
\begin{adjustbox}{width=\textwidth,center}
    \centering
    \begin{tabular}{|c|c|c|cc|c|}
    \noalign{\hrule height 1pt}
        \multirow{2}*{\textbf{Alternative}} & \multirow{2}*{\textbf{\#Number (\%)}} & \multirow{2}*{\textbf{Example App}} & \multicolumn{2}{c|}{\textbf{\underline {iOS Platform}}} & \textbf{\underline{Android Platform}}  \\ 
        
        ~ & ~ & ~ & \textbf{Example Model} & \textbf{Model Function} & \textbf{Function Reservation} \\ 
        
        \noalign{\hrule height 1pt}
        
        \multirow{3}*{Select other model} & \multirow{3}*{58 (25.66\%)} &
         Hair Color Changer & hair\_segmentation.tflite & hair segmentation & Yes \\ 
        
        & & Rock Identifier: Stone ID & object-detection-centernet.mlmodelc & object detection & Yes \\ 
        
        & & Hello Pal:Talk to the World & M\_SenseME\_Face\_Video\_5.3.3.model & face detection & Yes \\ 
        \noalign{\hrule height 1pt}
        \multirow{3}*{Remove functionality} & \multirow{3}*{55 (\%24.34\%)} &
        Strike: Bitcoin \& Payments & inference\_graph.tflite & object detection & No \\ 
        
        & & AfterShip Package Tracker & SSDOcr.mlmodelc & OCR text detection & No \\ 
        
        & & adidas CONFIRMED & FindFour.mlmodelc & object detection & No \\ 
        
        \noalign{\hrule height 1pt}
        \multirow{2}*{Share On-device model} & \multirow{2}*{45 (\%19.91\%)} & \multirow{2}*{TikTok} & tt\_face\_attribute\_age.model & age prediction & Yes \\ 
        
        & & & tt\_face.model & face detection & Yes \\
        
        \noalign{\hrule height 1pt}
        \multirow{2}*{Select On-cloud model} & \multirow{2}*{34 (15.04\%)} &
        Homebase Employee Scheduling & SSDOcr.mlmodelc & OCR text detection & Yes \\ 
        
        & & Postmates - Food Delivery & DocScanCSCModel.tflite & object detection & Yes \\

        \noalign{\hrule height 1pt}
        
        \multirow{3}*{Obfuscate/Encryption model} & \multirow{3}*{24 (10.62\%)}
        & \multirow{3}*{EnhanceFox - AI Photo Enhancer} & EVBodyML100S16FP16.mlmodelc & \multirow{3}*{photo enhancement} & Yes \\
        & ~ & & facesr\_380000\_pt.mlmodelc & ~ & Yes \\ 
        & ~ & & RealESRGANv2-animevideo.mlmodelc & ~ & Yes \\ 
        
         \noalign{\hrule height 1pt}
        Unknown & 10 (4.42\%) & TickPick: No Fee Tickets & SSDOcr.mlmodelc & OCR text detection & Yes \\ 
    
    \hline
    \end{tabular}
\end{adjustbox}
    \label{tab:notuse}
    \vspace{-0.5cm}
\end{table*}

\subsubsection{Results}
Table~\ref{tab:notuse} shows the investigation results of model sharing and replacement for 226 models in iOS-Android app pairs.
The column \emph{Alternative} represents the alternative method adopted by the corresponding Android app.
The \emph{unknown} in column \emph{Alternative} represents we cannot determine how this model is replaced now.
The column \emph{Number (\%)} represents the number and percentage of models are treated in this way.
The column \emph{Example App} shows some representative apps which are treated in this way. 
The column \emph{Example Models} and \emph{Model Functions} represent the detected on-device models in iOS apps and their functionalities.
The column \emph{Function Reservation} shows whether the identical function is reserved in the corresponding Android apps. 

58 (25.66\%) out of 226 models choose alternative on-device models in corresponding Android apps.
The app \emph{Hello Pal} in Table~\ref{tab:notuse} employs a third-party SDK from SenseTime for face detection on iOS, but the Android app leverages Google's face detection technology.
Because of the difference in platforms, the same app's can select more appropriate frameworks~\cite{nguyen2019machine}.
Similar to choosing on-cloud models, we discover developers, particularly those from small and medium-sized developing teams, tend to select DL frameworks like Core ML and TF Lite that are more practical, simple to use, and more affinity with mobile systems~\cite{chooseDL, findDL}.

In corresponding Android apps, 55 (24.34\%) of the 226 models are eliminated. 
We find that the functionalities of these eliminated models are not fundamental to the apps, so their removals have no impact on the apps' primary business.
For example, the \emph{inference\_graph.tflite} model in Table~\ref{tab:notuse}, which is used for inference in the \emph{Strike:Bitcoin \& Payments} app, is removed in the Android counterpart, but no effect for this app's business.
According to the survey report~\cite{diffTeam}, Android and iOS apps are frequently developed by different teams in one software business due to their varied technical routes.
Thus, Android versions may use various technical solutions, like as DL frameworks, resulting in small variations in the auxiliary functions of the same app on different platforms.

45 (19.91\%) of the 226 models are found to be shared between Android and iOS apps.
We notice that the functionalities of these re-used models are always customized by the apps' features and plays vital parts in their apps.
TikTok, for instance, reuses two on-device models: \emph{tt\_face\_attribute\_age.model}, which is used for age prediction, and \emph{tt\_face\.model}, which is used for face detection in both Android and iOS apps.
 
34 (15.04\%) out of 226 models are replaced with on-cloud DL models.
Different from Core ML on iOS, Google's ML Kit~\cite{mlKit} offers practical and convenient on-cloud DL model solutions. 
Android developers can activate on-cloud deep learning models for text, face, and barcode recognition with a few lines of code.
Note that differences in hardware and software configurations can significantly affect the performance of on-device models, particularly on the fragmented Android platform. IPhones generally have similar hardware and software configurations, but Android devices are plagued with fragmentation issues, causing compatibility concerns for app developers across various Android devices. Certain Android devices with suboptimal hardware conditions may severely impact the efficiency of on-device models, rendering them unsuitable for use. In such scenarios, on-cloud models are a more viable option, given that they have lower hardware and software requirements for phones and are better suited for the heavily fragmented Android platform. Additionally, while on-cloud models may have network requirements, they can be more cost-effective in terms of engineering efforts and deployment, thus making them a more appealing choice for some Android developers.
For example, as shown in Table~\ref{tab:notuse}, \emph{Homebase Employee Scheduling} and \emph{Postmates - Food Delivery} use on-cloud models provided by ML Kit to detect objects in Android apps to replace on-device models \emph{SSDOcr.mlmodelc} and \emph{DocScanCSCModel.tflite} in iOS apps.

24 (10.62\%) out of 226 models in the Android app received extra protections, including weights compression and model obfuscation.
It is tough for us to obtain extra details on the obscured model.
Android apps are widely perceived as being less secure than iOS apps.
Therefore, developers have increased security measures for Android apps.
To prevent the model from being readily taken or attacked,
\emph{EnhanceFox - AI Photo Enhancer} in Table~\ref{tab:notuse} obfuscates the DL models in Android apps, that could have a positive impact on user privacy and corporate assets.

10 (4.42\%) out of 226 models cannot currently be identified for replacement in Android apps.

\subsubsection{Suggestions}
\label{sec:suggestions}
We offer suggestions for both app developers and DL framework providers, which may support future development in this field.

To retain current DL framework users and appeal to a broader user base, framework providers should optimize the framework in two areas. 
First, to encourage the use of deep learning frameworks in multiple platforms, it is essential to make the cross-platform migration of the model more convenient and concise, while improving the cross-platform compatibility of the model. 
Our analysis reveals that despite the claims of deep learning frameworks to support both Android and iOS platforms, only a small percentage of models, i.e., 17.63\% out of 1,883 models, are shared between the two platforms.
Currently, the migration of a deep learning model from Android to iOS platforms involves converting the model from its original format to a format compatible with the iOS platform, while ensuring that the functionality and accuracy of the model remain unchanged. This conversion process may require modifications to be made to the model's architecture, code, and parameters, while considering the differences in hardware and software configurations between the two platforms.
Developers may find the engineering efforts involved in using the current framework's cross-platform migration model approach relatively high, as compared to using a more convenient approach specific to the current platform to achieve the same deep learning functionality, such as the on-cloud model. Therefore, the proportion of current developers reusing models between Android and iOS is not high.
Second, the ease of use of the framework should be improved by providing a comparable and accessible solution for utilizing deep learning functionalities on the iOS platform as on the Android platform, along with robust cross-platform support.
Deep learning features, such as text recognition, QR code recognition, face recognition, and object recognition, are more accessible to developers in Android apps compared to iOS. 
Due to cost-saving advantages, Android developers are more inclined to use pre-existing deep learning techniques provided by frameworks rather than reuse their own models.
In cases where on-device models are reused between the two platforms, they are often customized by app developers to provide tailored functionalities, which cannot be provided by DL frameworks.
In conclusion, cross-platform compatibility and ease of use should become the key direction of future framework optimization.

From the developer's perspective, it is essential to understand that differences in hardware and software between Android and iOS platforms can significantly affect the accuracy and efficiency of the model. Furthermore, the availability of specific hardware components, such as GPUs or accelerometers, can vary between the two platforms, influencing model selection and usage. Therefore, developers should carefully consider these factors and choose a more suitable framework when developing mirror apps on different platforms.
Moreover, developers should prioritize the cross-platform compatibility of app-customized models by pre-defining cross-platform interfaces and security safeguards for these models. This can mitigate the engineering efforts and security concerns associated with developing mirror apps for other platforms in the future. In summary, selecting a more appropriate DL framework for the app's target users and designing interfaces and security for custom models in advance can ease the developer's engineering efforts and better maintain the consistency of one app on different platforms.

\begin{summary}[]
Due to the different ecosystems of the Android and iOS platforms, only a tiny number of Android apps fully reuse the on-device models of iOS apps. 
25.66\% of iOS apps' on-device models are replaced in Android by selecting more convenient models.
17.63\% of on-device models, which provide customized functionalities for apps, are shared by apps across Android and iOS.
Existing DL framework vendors should simplify developers' use of fundamental DL models and improve cross-platform interoperability and security for developers' on-device models to acquire market share.
\end{summary}


\section{RQ3: How robust are on-device models on iOS against adversarial attacks?}

\subsection{Motivation}
RQ1 demonstrates that on-device models are widely used and serve fundamental roles in iOS apps.
We discover in RQ2 that while most apps will not utilise the exact same on-device model for iOS and Android, the model that serves as the app's basic functionality will be shared between the two platforms. 
Most of the shared models are exposed directly in the source files, and only a very few developers choose to provide additional protection to the models.
Shared models include the white-box model and the gray-box model.

White-box on-device models on Android are proved to be vulnerable to adversarial attacks~\cite{huang2021robustness, huang2022smart, karim2020adversarial}.
Due to the closed-source ecosystem and the inability of current methods to attack the gray-box model~\cite{vivek2018gray, ebrahimi2017hotflip}, attacking iOS apps is considered more challenging than attacking Android apps~\cite{garg2021comparative, dehling2015exploring, fredrikson2015model}, resulting in the lack of methods that focus on exploiting iOS-specific on-device models.
However, some studies have demonstrated that there are concurrent cross-platform issues in iOS and Android apps~\cite{ahmad2013comparison, gronli2014mobile, aljedaani2019comparison, garg2021comparative}.
Regardless of the fact that investigating iOS security issues is regarded as being more challenging, given the widespread use of on-device models on iOS, we study how robust these models are against adversarial attacks.

In this RQ, we first present our methodology of attacking on-device models in Section~\ref{sec:attackPipe}, including white-box and iOS-specific gray-box models.
Then, we carry out experiments to evaluate and compare the effectiveness of our approach in Section~\ref{sec:atkEva}.
Finally, we successfully attacked the relevant functionality in real-world iOS apps by taking advantage of the flaws of the on-device models in Section~\ref{sec:realworld}.

\subsection{Methodology of Model Attacking}
\label{sec:attackPipe}

\begin{figure*}[htbp]
\vspace{-0.5cm}
\setlength{\abovecaptionskip}{10pt} 
\setlength{\belowcaptionskip}{10pt}
    \centering
    \includegraphics[width=0.9\textwidth]{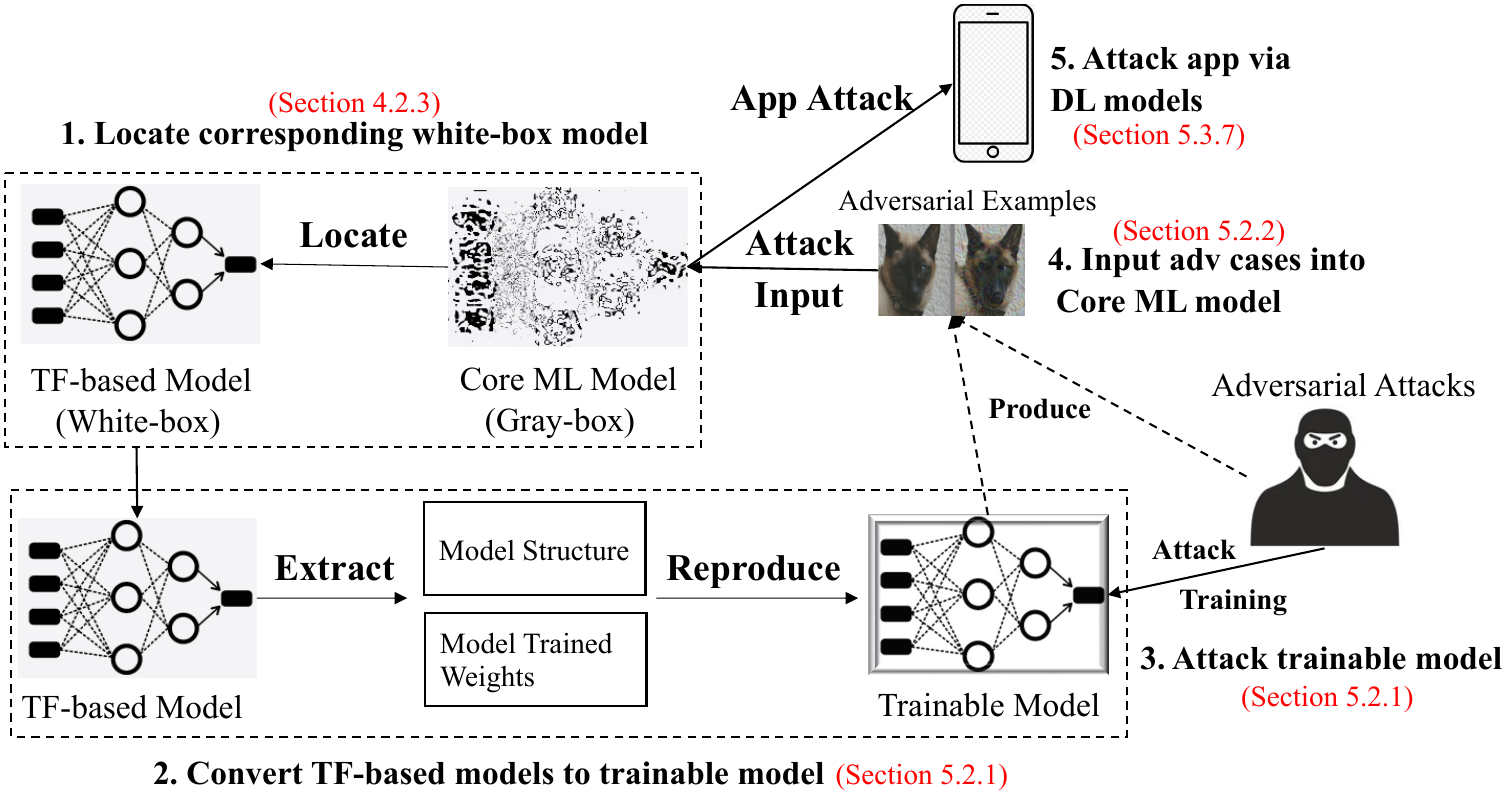}
    \caption{The Methodology of Model Attacking}
    \label{fig:rq3}
\vspace{-0.5cm}
\end{figure*}

According to the empirical study in Section~\ref{sec:frameCompare}, there are two categories of on-device models inside current iOS apps:
(1) White-box Model: both the trained network weights and model structure can be extracted, for example, TF Lite and TensorFlow models; (2) Gray-box Model: only the model structure can be obtained (Core ML models).
Figure~\ref{fig:rq3} shows the methodology of our model attacking.
Regarding white-box models, we propose a more general attack approach for models of TF Lite and TensorFlow in this study.
Current approaches perform poorly when attacking gray-box on-device models of Core ML~\cite{huang2021robustness, huang2022smart, karim2020adversarial, rauber2020fast}.
Regarding gray-box models, we use the Core ML models' white-box counterpart as a bridge to propose an more efficient pipeline for attacking Core ML models.
Our approach is also applicable to white-box and gray-box models that do not fit into these three frameworks.

\subsubsection{How to Attack White-box Models?}
\label{sec:attackWhite}
The attacking approaches proposed by Huang et al.~\cite{huang2021robustness, huang2022smart} require identifying the on-device model's pre-trained model for producing adversarial examples, and then using these adversarial examples to attack the on-device model.
Such approaches can only be used for fine-tuning models and requires the identification of the corresponding pre-trained model.
The fundamental reason why TF-based models cannot be directly attacked is that it is hard to continue training the models in these formats and unable to update the parameters in the model by backpropagation.
Therefore, Huang et al.~\cite{huang2021robustness, huang2022smart} choose to first identify the pre-trained model from TensorFlowHub~\cite{tfhub} which could perform backpropagation to generate adversarial examples.

\begin{figure*}[htbp]
\setlength{\abovecaptionskip}{10pt} 
\setlength{\belowcaptionskip}{10pt}
    \centering
    \includegraphics[width=\textwidth]{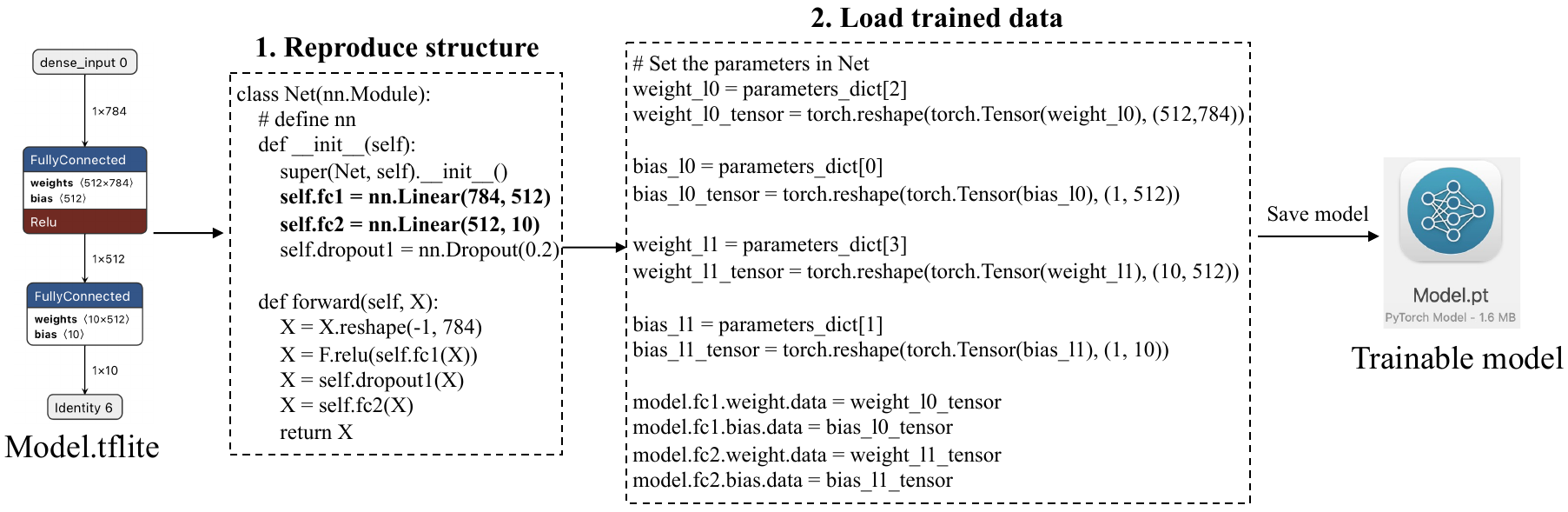}
    \caption{A concrete example of model conversion}
    \label{fig:conversionExp}
    \vspace{-0.5cm}
\end{figure*}

Our approach does not rely on pre-trained models to generate adversarial examples and also supports direct attacks on models without fine-tuning.
Step 2 (\emph{Convert TF-based models to trainable model}) in Figure~\ref{fig:rq3} shows the pipeline of this approach.
In our approach, given an on-device model, we first retrieve its parameters and structure information and then use them to reproduce the model in a trainable format, such as the \emph{.pt} from PyTorch.
Expressly, we feed the extracted parameters to our reconstructed model structure layer by layer.
After converting to a trainable model, as shown in step 3 (\emph{Attack trainable model}) in Figure~\ref{fig:rq3}, we employ adversarial attacks on the reproduced trainable model to generate adversarial examples, which will be used to attack the original on-device model in the corresponding app.
Figure~\ref{fig:conversionExp} shows a concrete example of converting an on-device model into a trainable model format, namely\emph{pt} of PyTorch.
We first reproduce all the neural network structures of this model. Then we load the trained weight dictionary (\emph{parameter\_dict}) of the original on-device model, and feed the weights of the original model for our reproduced network structure according to the shape and name of each network layer. Finally, we save the replicated model in trainable format (\emph{Model.pt}).
We generate adversarial examples based on the training of our replicated trainable models (\emph{Model.pt}). 
we then use these adversarial examples to attack the original on-device model (\emph{Model.tflite}).

\subsubsection{How to Attack Gray-box Models?}
On-device models of the Core ML framework do not share trained parameters with third parties and may compress or encrypt model parameters~\cite{modelEncry}, resulting in the inability to retrieve the parameters of the trained model.
This type of gray-box model is tough to be successfully attacked by current approaches without knowing trained parameters.
Steps 1, 2, 3 and 4 in Figure~\ref{fig:rq3} demonstrate our attack pipeline for gray-box models.
In Section~\ref{sec:modelChange}, we find that some iOS and Android apps share on-device models and that some on-device models of Core ML framework are converted from models in Android apps~\cite{CoreMLConvert}.
Therefore, following steps in Section~\ref{sec:ps2}, we first locate the matching on-device model before conversion in the corresponding Android app (Step 1 in Figure~\ref{fig:rq3}).
We retrieve the model's parameters and reconstruct it in a trainable format (Step 2 in Figure~\ref{fig:rq3}). 
Then, we attack the reconstructed model of its counterpart to generate adversarial examples for targeting the original on-device model of Core ML (Step 3 and 4 in Figure~\ref{fig:rq3}).

\subsection{Evaluation of Attacking}
\label{sec:atkEva}
The effectiveness of our proposed approach to attack the white-box model is evaluated first, followed by the effectiveness of our proposed method to attack the gray-box model.

\subsubsection{Dataset}
\begin{table*}[htbp]
  	\vspace{-0.8cm}
\setlength{\abovecaptionskip}{0pt} 
\setlength{\belowcaptionskip}{10pt}
  \centering
  \caption{Details of the selected 10 models}
  \begin{adjustbox}{width=\textwidth,center}
    \begin{tabular}{|c|c|c|c|c|}
    \noalign{\hrule height 1pt}
    \textbf{ID} & \textbf{App Name} &\textbf{Model in iOS} & \textbf{Model in Android} & \textbf{Model Function} \\
    \noalign{\hrule height 1pt}
    1     & hp smart & doc\_classification.mlmodelc & doc\_classification.tflite & document classification \\
    2     & paypal - send, shop, manage & ObamModel.mlmodelc & obamModel.tflite & image classification \\
    3     & merlin bird id by cornell lab & geo\_v18.mlmodelc & geo\_v17.tflite & bird classification   \\
    4     & seek by inaturalist  & optimized\_model.mlmodelc & optimized\_model.tflite & identify plants and animals in pictures\\
    5     & smart bird id & NABirdsImageClassifier.mlmodelc & BirdImageClassifier.tflite & identify birds in pictures\\
    6     & sticker maker studio & deeplabv3\_mnv2\_pascal\_trainval.mlmodelc & deeplabv3\_mnv2\_pascal\_trainval.tflite & image segmentation \\
    7     & scentbird & FindFour.mlmodelc & findfour.tflite & detect box in pictures \\
    8     & scentbird & FourRecognize.mlmodelc & fourrecognize.tflite & detect box in pictures\\
    9     & gradient: celebrity look alike & gender\_nn.mlmodelc & gender\_nn.tflite & identify gender in pictures\\
    10    & Hoop - Make new friends & Nudity.mlmodelc & optimized\_nudity\_graph.tflite & identify nudity in pictures\\
    \noalign{\hrule height 1pt}
    \end{tabular}%
    \end{adjustbox}
  \label{tab:dataset}%
  	\vspace{-0.2cm}
\end{table*}%

We evaluate the performance of our proposed attack pipeline in real-world industrial iOS apps from the Apple App Store.
We select 10 representative on-device models of the Core ML framework for testing our approach.
These 10 models all have clear usage scenarios in iOS apps and serve core functions.
Table~\ref{tab:dataset} shows the detail of selected 10 on-device models.
The column \emph{App Name} represents the apps to which the on-device models belong. 
The column \emph{Model in iOS} represents the gray-box model in iOS apps.
The column \emph{Model in Android} represents gray-box models' located white-box counterparts in the Android app.
The column \emph{Model Function} represents the identified model function in apps.
In our present analysis of commonly utilized apps, we have only identified the reuse of image-based on-device models between Android and iOS platforms. 
Therefore, the selected models are all image-based.
However, the advent of large language models (LLMs) suggests the potential for expansion of this phenomenon to other domain models such as natural language processing (NLP), speech recognition, and more. 
As of now, these models are not yet prevalent in cross-platform reuse. 
Nevertheless, our methodology exhibits a high degree of flexibility and adaptability, making it well-suited to accommodate such future developments. 
We will direct our research efforts toward exploring the security implications of these emergent domains and their corresponding large models in the future.

\subsubsection{Baseline}
To demonstrate the effectiveness of our newly proposed method to attack the white-box model, we use the ModelAttacker proposed by Huang et al.~\cite{huang2021robustness, huang2022smart} as the baseline to carry out a control experiment.
ModelAttacker hacks DL models using adversarial attacks by first identifying highly similar pre-trained models from TensorFlow Hub~\cite{tensorflowHub}.
Then, it attacks the identified pre-train model to produce adversarial examples for the original model.

\subsubsection{Experimental Setup}
We first identify Android white-box counterparts of Core ML models (as shown in column \emph{Model in Android}).
Second, following steps in Section~\ref{sec:attackWhite}, we convert these on-device models into trainable models (\emph{.pt}).
We find 10 random images for each model in Table~\ref{tab:dataset} by referring to their identified functionalities as the original inputs~\cite{huang2021robustness, huang2022smart}.
To ensure the validity of the input images selected for our experiments, we employ a two-step collection process. 
In the first step, we identify target images matching the function of the model and the app's specific usage scenario. 
For instance, for model 5 (\emph{NABirdsImageClassifier.mlmodelc}) in Table~\ref{tab:dataset}, which is used for bird identification, we selected various bird images as input for the model. 
In order to ensure the diversity of the input test set, we will deliberately select images that encompass distinct object categories, backgrounds, and lighting conditions. This approach will ensure that our test set is a true reflection of the variability encountered in real-world images, and will help to establish the efficacy of our method in practical settings.
In the second step, we manually input the selected experimental images directly into the corresponding application scenes of the model in the app to verify the model's correct recognition of these images. Following this process, we obtained a set of valid input images. 

In this study, we follow the approach of Huang et al.~\cite{huang2022smart, huang2021robustness} and randomly select 20 images for each on-device model. To ensure the diversity of the selected images, we partition the 20 images into two subsets such that each subset consists of dissimilar images.
To quantitatively evaluate the diversity of the new dataset, we use Maximum Mean Discrepancy (MMD)~\cite{borgwardt2006integrating} as a metric. MMD is a measure of the discrepancy between two probability distributions, frequently used in machine learning to compare the similarity between datasets. MMD compares the means of the two datasets in a reproducing kernel Hilbert space (RKHS) and quantifies their differences. Higher MMD values indicate greater dissimilarity between two datasets or samples.
Based on prior research~\cite{dziugaite2015training, yan2017mind}, we require that the MMD values of the two datasets exceed 0.6 to ensure their diversity. If the MMD values fall below this threshold, we re-collect images until the requirements are satisfied.

Finally, we use selected images as input to attack the converted trainable models to generate adversarial examples and use generated adversarial examples to evaluate if they will be misclassified by original on-device models on Android and iOS.

In the control experiment, we first find the most similar pre-train models from TensorFlowHub for each model in the column \emph{Model in Android} by following the experiment setup of the baseline.
Then, we use selected 10 images as input to attack the pre-train models to generate adversarial examples.
Finally, we use the generated adversarial examples to attack the original iOS on-device models.
The baseline requires locating pre-trained models with greater than 80\% structural similarity~\cite{huang2021robustness}.
However, we cannot find such models on TensorFlowHub for model 9 (\emph{gender\_nn.tflite}) and model 10 (\emph{optimized\_nudity\_graph.tflite}).
So when comparing the attack success rate of the baseline, we only consider the success rate of  models 1 to 8.

\subsubsection{Evaluation Metrics}
Successful adversarial examples must meet two requirements:  (1) make the model misclassify (2) changes made in the original image of the input cannot be noticed by humans.
Following the same evaluation steps with related works~\cite{huang2021robustness, huang2022smart, karim2020adversarial}, we invite three PhD students with experience in adversarial attacks to manually evaluate whether the modifications made to adversarial example images are too subtle to be noticed.
The attack is deemed successful if the three volunteers think the modifications to the image cannot be easily recognised.
Otherwise, the attack is considered failed.
we count the number of examples that successfully misclassify the model among these 10 input adversarial examples and count the success rate of different types of adversarial attacks.

\subsubsection{Adversarial Attack Algorithms}

The whole process of adversarial attack can be summarized as
\begin{equation}
    x^{'}_{adv} = x + \epsilon * attack_i(\nabla J (\theta,\ x,\ y_{true}))
\end{equation}
where $y_{true}$ represents the original label or class of input image $x$, $\epsilon$ represents a multiplier to ensure the perturbations between input images and adversarial images are small and its value is empirically determined in the experiments to produce unnoticeable image perturbations, $attack_i$ represents the $i_{th}$ type of adversarial attack, $\theta$ represents parameters, and $J$ is the loss.

We select 10 representative attacks: Pointwise (PW) Attack~\cite{yang2019adversarial}, Boundary (BD) Attack~\cite{croce2020minimally}, DDN Attack~\cite{rony2019decoupling}, DeepFool (DF) Attack~\cite{moosavi2016deepfool}, FGSM Attack~\cite{goodfellow2014explaining}, BIM~\cite{kurakin2018adversarial}, PGD~\cite{he2019towards}, C\&W Attack~\cite{carlini2017towards}, Newton Fool (NF) Attack~\cite{jang2017objective} and Clipping-Aware Noise (CAN) Attack~\cite{rauber2020fast} as all these attacks have been proved effective in many tasks and widely-used to evaluate the robustness of on-device models~\cite{huang2021robustness, huang2022smart, karim2020adversarial}.

\subsubsection{Results}
\begin{table*}[htbp]
\vspace{-0.8cm}
\setlength{\abovecaptionskip}{0pt} 
\setlength{\belowcaptionskip}{10pt}
  \centering
  \caption{Attack results of Core ML models on iOS and their counterparts on Android}
    \begin{adjustbox}{width=\textwidth, center}
    \begin{tabular}{|cc|ccc|ccc|ccc|ccc|ccc|ccc|ccc|ccc|ccc|ccc|ccc|}
    \noalign{\hrule height 1pt}
    \textbf{ATK} & \textbf{Ep} & \multicolumn{3}{c|}{\textbf{M1}} & \multicolumn{3}{c|}{\textbf{M2}} & \multicolumn{3}{c|}{\textbf{M3}} & \multicolumn{3}{c|}{\textbf{M4}} & \multicolumn{3}{c|}{\textbf{M5}} & \multicolumn{3}{c|}{\textbf{M6}} & \multicolumn{3}{c|}{\textbf{M7}} & \multicolumn{3}{c|}{\textbf{M8}} & \multicolumn{3}{c|}{\textbf{M9}} & \multicolumn{3}{c|}{\textbf{M10}} & \multicolumn{3}{c|}{\textbf{Ave (ATKs)}} \\
    \noalign{\hrule height 1pt}
     & & \textbf{M} & \textbf{A} & \textbf{I} &  \textbf{M} &\textbf{A} & \textbf{I} &  \textbf{M}&  \textbf{A} & \textbf{I} & \textbf{M}&  \textbf{A} & \textbf{I} & \textbf{M}&  \textbf{A} & \textbf{I} & \textbf{M}&  \textbf{A} & \textbf{I} & \textbf{M}&  \textbf{A} & \textbf{I} & \textbf{M}&  \textbf{A} & \textbf{I} & \textbf{M}&  \textbf{A} & \textbf{I} & \textbf{M}&  \textbf{A} & \textbf{I} & \textbf{M}&  \textbf{A} & \textbf{I} \\
    
    \hline
    
    PW & 3.5  & 0.2  & 0.6  & {0.5}   & 0.4  & {0.7}  & {0.7}   & 0.3  & {0.6}  & {0.6}  & 0.5   & 0.7  & 0.7   & 0.6  & 0.8  & 0.7   & 0.4  & {0.8}  & {0.8}    & 0.6 & 0.8  & 0.8    & 0.3 & 0.6  & 0.6    & - & 0.8  & {0.7}    & - & 0.8  & 0.7    & 0.41 & {0.72}  & {0.68} \\
    
    BD & 5.5    & 0.3 & {0.6}  & {0.6}   & {0.4} & 0.7  & 0.7   & 0.4 & 0.8  & 0.8   & 0.5 & 0.7  & 0.7   & 0.2 & {0.7}  & {0.7}  & 0.6 & 0.8  & 0.8   & 0.3 & {0.6}  & 0.5   & 0.3 & {0.5}  & {0.5}  & - & 0.8  & 0.8   & - & 0.8  & 0.8   & {0.28} & {0.70}  & {0.69}\\
    
    DDN & 0.8  & 0.4 & 0.8 & {0.7}  & 0.1 & {0.6} & {0.6}    & 0.2 & 0.7 & 0.6    & 0.2 & 0.7 & 0.7   & 0.3 & 0.6 & 0.6   & 0.3 & {0.8} & {0.8}  & 0.2 & 0.7 & 0.7   & {0.2} & 0.7 & 0.7    & - & 0.8 & 0.8   & - & 0.7 & 0.5  & {0.24} & {0.71}  & {0.67}\\
    
    DF & 1.7  & 0.1 & {0.6} & {0.6}  & 0.2 & 0.6 & 0.6    & 0.4 & {0.7} & {0.7}    & 0.3 & {0.8} & {0.8}   & 0.3 & 0.8 & 0.8   & 0.2 & {0.7} & {0.7}  & 0.3 & {0.5} & {0.5}    & 0.3 & 0.7 & 0.7    & - & 0.9 & 0.9   & - & {0.6} & {0.6}  & 0.26 & {0.69}  & {0.69} \\
    
    FGSM  & 0.02 & 0.3 & 0.6 & 0.6  & 0.4 & {0.6} & {0.6}   & 0.5 & 0.8 & 0.8    & 0.3 & 0.7 & 0.7   & 0.4 & 0.7 & 0.7   & 0.2 & {0.8} & {0.8}  & 0.3 & {0.7} & {0.7}    & 0.3 & 0.7  & 0.7    & - & 0.8 & 0.8   & - & 0.8 & 0.7    & 0.34 & {0.72}  & {0.71}\\
    
    BIM & 2.5    & 0.3 & {0.7} & {0.7}   & 0.2 & {0.7} & {0.7}    & 0.5 & 0.9 & 0.9    & 0.2 & {0.7} & {0.7}   & 0.1 & {0.6} & {0.6}   & 0.6 & {0.7} & {0.7}  & 0.3 & 0.7 & 0.7    & 0.4 & {0.7} & {0.7}    & - & 0.9 & 0.9   & - & 0.8 & 0.6   & 0.35 & {0.70}  & {0.68}\\
    
    PGD & 8   & 0.3 & {0.6} & {0.5}  & 0.2 & {0.6} & {0.5}    & 0.6 & 0.9 & 0.8    & 0.5 & {0.7} & {0.7}    & 0.2 & {0.6} & 0.5   & 0.1 & 0.5 & 0.5  & 0.3 & 0.7 & 0.7    & 0.4 & 0.8 & 0.8    & - & {0.7} & {0.7}   & - & 0.7 & 0.7  & 0.33 & {0.68}  & {0.67}\\
    
    C\&W & 0.2 & 0.5 & 0.7 & 0.7  & 0.2 & 0.6 & 0.6    & 0.1 & {0.8} & {0.8}    & 0.6 & 0.8 & 0.8   & 0.4 & {0.8} & {0.7}   & 0.4 & 0.7 & 0.7  & 0.2 & 0.5 & 0.5    & 0.2 & 0.6 & 0.6    & - & 0.9 & 0.9   & - & {0.7} & {0.7}    & 0.36 & {0.71}  & {0.70}\\

    NF & 9.5   & 0.4 & {0.7} & {0.7} &   0.3 & 0.7 & 0.7    & 0.3 & 0.7 & 0.7    & 0.4 & 0.7 & 0.6  & 0.5 & {0.8} & 0.8   & 0.4 & 0.7 & {0.6}  & 0.2 & {0.7} & {0.7}    & 0.3 & {0.6} & {0.6}    & - & 0.8 & 0.8   & - & 0.7 & 0.7 &   0.35 & {0.71}  & {0.69}\\
    
    CAN & 20   & 0.5 & 0.8 & 0.8  & 0.4 & 0.9 & 0.9    & {0.5} & 0.9 & 0.9    & 0.4 & 0.8 & 0.8   & 0.6 & 0.9 & 0.9    & 0.5 & 0.8 & {0.7}  & 0.5 & 0.8 & 0.8    & 0.6 & 0.9 & 0.9    & - & 0.8 & 0.8   & - & 0.9 & 0.9 & {0.49} & \textbf{0.85}  & {\textbf{0.84}}\\
    \noalign{\hrule height 1pt}
    Ave &   ~   & 0.33 & {0.67} & {0.64}  & {0.28} & {0.67} & {0.66}    & {0.38} & {0.78} & {0.76}    & 0.39 & {0.73} & {0.72}   & 0.36 & {0.73} & {0.70}  & 0.37 & {0.73} & {0.72}  & 0.34 & {0.67} & {0.66}     & {0.33} & {0.68} & {0.68}   & - & \textbf{0.82} & {\textbf{0.81}}   & - & {0.76} & {0.70} & {0.34} & {\textbf{0.72}}  & {0.70} \\
    \noalign{\hrule height 1pt}
    \end{tabular}%
    \end{adjustbox}
  \label{tab:results3}%
  \vspace{-0.3cm}
\end{table*}%

Table~\ref{tab:results3} shows the adversarial attack results in the dataset.
The column \emph{ATK} shows the type of adversarial attacks.
The column \emph{Ep} denotes the empirical value of the $\epsilon$ that provides the best performance of the current attack in the experiment.
The columns from \emph{M1} to \emph{M10} represent models with ids 1 to 10 in Table~\ref{tab:dataset}.
The subcolumns \emph{M} represent the attack results of the baseline ModelAttacker.
The subcolumns \emph{A} and \emph{I} represent the attack results of our approach on Android and iOS models, respectively.

As shown in the column \emph{Ave (ATKs)}, the average success rate of ModelAttacker's attack is 0.34, which is much lower than that of our approach 0.72 on white-box models and 0.70 on gray-box Core ML models.
The baseline for producing adversarial examples by attacking the corresponding pre-training model has limits and is only applicable for fine-tuning on-device models.
Our proposed attack method is demonstrably more effective and is applicable to non-pre-trained models like \emph{M9} and \emph{M10}.

As shown in the row \emph{Ave}, all of these 10 Core ML models are effectively attacked by our approach, with a success rate of at least 64\%.
The average success rate of attacks against Android equivalents is 0.72 (shown in column \emph{Ave(ATKs)}, row \emph{Ave}), which is slightly higher than the success rate of attacks against the original Core ML models, which is 0.70.
The extremely close success rate implies that gray-box Core ML models are vulnerable to adversarial examples generated by targeting their white-box counterparts, demonstrating the effectiveness of our cross-platform attack approach.

We investigate why there is a slight distinction in the success rates of attacking Android and iOS models.
There are two reasons: (1) Different frameworks may optimise the model differently. Consequently, the models may be updated slightly during the framework conversion process, resulting in subtle variations in model effects.
(2) Android app and iOS app version updates are not synchronised, so the model versions inside the app are not consistent, such as models \emph{geo\_v18.mlmodelc} and \emph{geo\_v17.tflite} in app \emph{merlin bird id by cornell lab}.
Different versions of the model may have a slightly different effect.

The model \emph{M9} (\emph{gender\_nn.mlmodelc}) is the most vulnerable with the highest average success rate of is 81\%,  since \emph{gender\_nn.mlmodelc} has the simplest structure, consisting of several residual blocks.
The Clipping-Aware Noise Attack has the most significant attack effect (85\% average success rate), whereas the success rates of the other 9 attacks are much lower (68\% - 72\%).
We find that the Clipping-Aware Noise Attack generates adversarial images with a narrower range of perturbations that are more likely to survive manual recognition and so have a higher success rate.

\subsubsection{Results Discussion}

Our experiments conclusively demonstrate that the black-box models present in iOS apps, which can match white-box counterparts, are inherently insecure. 
As outlined in Section~\ref{sec:ratio}, we discover that 332 (17.63\%) out of the total 1883 models are shared across both iOS and Android platforms and that these models are frequently reused in critical functionalities of the app. 
This finding highlights a significant ratio that cannot be ignored. 
Given the insecurities present in these models, both academia and industry must pay close attention to the potential unforeseen consequences and take necessary steps to prevent them.

In recent studies, researchers have analyzed the robustness of on-device models in Android apps against adversarial attacks. Our findings indicate that our proposed approach of converting the on-device model into a trainable format and generating adversarial examples through its own training is more effective than common adversarial attacks or attacking the corresponding pre-trained model before attacking the on-device model. Our approach can increase the success rate of attacks from about 45\% to 76\% when attacking white-box models.
In the work of Deng et al.~\cite{deng2022understanding} and Huang et al.~\cite{huang2021robustness}, C\&W and inversion attacks are found to be the most effective countermeasures against white-box models. However, we find that the Clipping-Aware Noise (CAN) attack, which has been recently proposed, has the highest success rate of the attack so far. Compared to C\&W and inversion attacks, CAN attack is relatively new and thus should receive more attention in terms of prevention.

In our black-box model attack experiments, we observed that compared to the NES algorithm~\cite{ilyas2018black} and ModelAttacker, our approach demonstrates a higher success rate in attacking partial black-box models with potential security threats.
Despite the comparatively closed and secure nature of the iOS system, we have shown that on-device models in Android apps are vulnerable to similar threats as those present in iOS apps. To mitigate this issue, app developers should apply the same model protection measures used in the Android platform, such as model obfuscation and encryption, to models in the iOS platform.
Furthermore, DL framework providers could enhance the security of models and attract more users by providing a cross-platform security mechanism for models that use their framework.

In above related studies and our work, it is worth noting that the on-device models examined in our experiments primarily focus on image-related domains and hardly cover other domains, such as natural language processing (NLP) and speech recognition. This is due to the fact that on-device models related to computer vision are currently the most prevalent type of DL models utilized by real-world applications, as discussed in Section~\ref{sec:modelFunction}. 
The models identified in our experiments that are reused in both platforms are primarily image-related. 
From the perspective of attack methods, investigating a broader range of model types would provide a more comprehensive understanding of how different types of models perform in the face of adversarial attacks. Thus, in future work, we plan to investigate the security vulnerabilities of all types of models against adversarial attacks.


\subsubsection{Examples of Attacking Real-world iOS Apps}
\label{sec:realworld}
\begin{figure*}[htbp]
\vspace{-0.8cm}
\setlength{\abovecaptionskip}{0pt} 
\setlength{\belowcaptionskip}{10pt}
	\centering
	\subfigure[Attacking the feature of gender classification in the app \emph{gradient: celebrity look alike}.]{
		\begin{minipage}[t]{0.49\linewidth}
			\centering
			\includegraphics[width=0.8\linewidth]{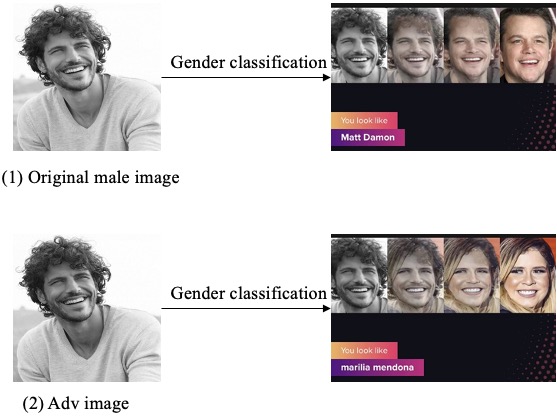}\\
			\label{fig:phAt1}
		\end{minipage}%
	}%
	\subfigure[Attacking the feature of bird classification in the app \emph{smart bird id}.]{
		\begin{minipage}[t]{0.49\linewidth}
			\centering
			\includegraphics[width=0.8\linewidth]{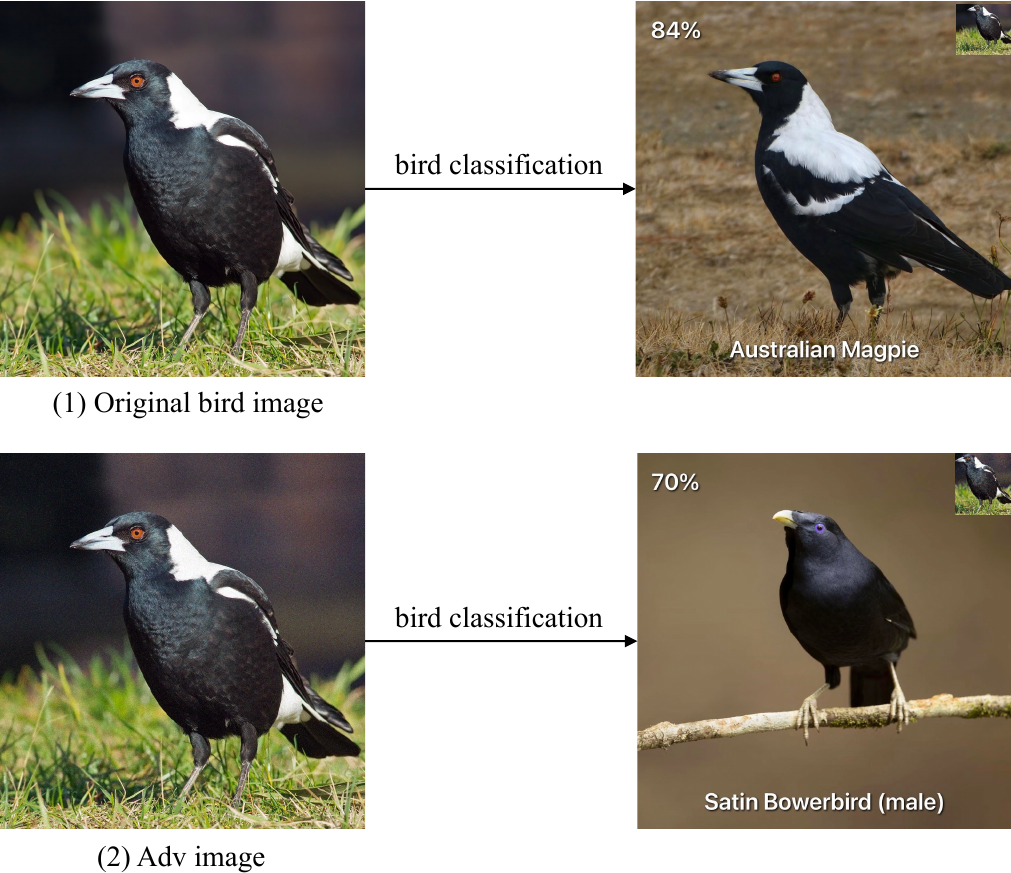}\\
			\label{fig:phAt2}
		\end{minipage}%
	}%
	\caption{Example of attacking real-world apps via DL models vulnerabilities}
	\label{fig:appComparision}
	\vspace{-0.8cm}
\end{figure*}

After locating the usage scenarios within iOS apps, we input the generated adversarial examples to the iOS apps to attack real-world iOS apps, as illustrated in the step 5 of Figure~\ref{fig:rq3}.
Figure~\ref{fig:phAt1} shows the attack results of the app \emph{gradient: celebrity look alike}.
The adversarial example image directly cause the app to misclassify the gender of the image from male to female.
Figure~\ref{fig:phAt2} shows the attack results of the app \emph{smart bird id}.
The adversarial example image cause the app to misclassify the bird type of the image from \emph{Australia Magpie} to \emph{Satin Bowerbird}.

\begin{summary}
Due to the reuse of models between platforms and apps, gray-box models of iOS-specific Core ML are also vulnerable to adversarial attacks via our approach.
The discovered vulnerabilities of on-device models inside iOS apps could disable some functionalities of real-world iOS apps.
Our findings reveal that the supposedly more secure iOS platform also has potential model security concerns, driving developers and DL frameworks to develop cross-platform model security defences against the possibility of cross-platform attacks.
\end{summary}

\section{Discussion}
In this section, we discuss some implications following our previous observation and experiments.

\subsection{Implication on Employing and Deploying On-device Models on iOS}

\subsubsection{Optimize NLP On-device Model Deployment.} In RQ1, most on-device models perform CV-related tasks, while the NLP-related on-device models are less common. This can be attributed to: (1) traditional/classical NLP models like LSTM and RNN are less efficient than Large Language Models (LLMs), however the development of LLMs for edge devices is hard. (2) NLP models inevitably come with dictionaries/word embedding layers, which are heavy and not easy to optimize. (3) Employing NLP models on edge devices should address the language challenge as a DL app may provide services to people from different countries.
Hence, this can be a future research direction as NLP is another crucial part for on-device deep learning models.

\subsubsection{Optimize Between-app On-device Model Development}
For developing a DL app in iOS and Android platform, DL framework vendors should optimize their products to facilitate the cross-platform on-device model development. Based on our findings in RQ2, potential solutions can be: (1) Implementing platform-specific optimisation to ensure that a well-trained model can be adapted for proprietary use on several platforms.
(2) offering unified APIs makes the development of on-device models for different platforms indistinguishable. (3) enhancing framework compatibility for various platforms.

\subsubsection{Optimize In-app On-device Model Collaboration}

In RQ1, we explore \emph{How many on-device models are there in one app?} in Section~\ref{sec:appNum} and \emph{What is the size of on-device models?} in Section~\ref{sec:modelSize}.
Current DL frameworks generally correspond to a single model for a single feature, and extensive usage of deep learning in iOS apps has resulted in an excessive and bloated on-device models for some apps.
Appropriately improving the management and scheduling of models can significantly enhance the app's utilisation of models.
A vital optimization direction is to optimize the management for all on-device models in an app.

From the DL framework vendor side, many current downstream tasks are based on the same type of pretrain models~\cite{huang2021robustness, tensorflowHub}.
Frameworks can attempt to slice and dice models and extract sliced model for usage in numerous tasks. 
Frameworks dynamically combine shared pretrain and fine-tuning model slices when on-device models are used. 
This increases the security of models by making it more difficult for an attacker to locate specific models. 
On the other hand, it can reduce the number and size of models within the app and optimise model utilisation efficiency.


\subsubsection{Prevent App Users from Locating to a Specific On-device Model using App Semantics}
We discover in RQ1 that locating the scenario of the model's usage in the app by the model's functionality is typically straightforward.
Although it is difficult to directly attack an iOS app, after discovering the model's usage scenario in the app, it is possible to use the model's vulnerability to disable the related functionality in order to further attack the app.
To prevent possible security issues, app developers should avoid leaving semantics that can be used to infer connections between models and functionalities.

\subsection{Implication for the Security of On-devices Models}

\subsubsection{Cross-platform Protection for Your Own On-device Models}
Although there are differences between the Android and iOS platforms, our research demonstrates that the same security vulnerabilities exist in the on-device model on both platforms.
In RQ3, we generated adversarial examples using the on-device model on the Android app and experimentally verified that these adversarial examples are equally valid for the Core ML model in iOS.
It is also possible to directly deactivate a portion of the iOS app's functionality.

From the app developer side, app developers should provide uniform protections when deploying on-device models in multiple platforms.
Also, the unified management of all on-device models within an app facilitates the reuse and deployment of uniform security measures for these models.

\subsubsection{Obfuscate On-device Models}
It is difficult to attack obfuscated on-device models with current ways of attacking on-device models.
However, we found in RQ2 that the percentage of obfuscated on-device models in current iOS apps is low (10.63\%).
Current iOS app developers and deep learning frameworks can design more methods to obfuscate the model to increase the model's security.



\subsubsection{Authenticate the Users of On-device Models}
Regarding DL app developers, the utilization of their on-device models can be secured by mandating user authentication and the provision of a unique token. From the perspective of model providers, it is essential to establish secure license distribution mechanisms from the server side and implement effective license management procedures on the client side prior to the utilization of their models in order to ensure security.

\section{Threats to Validity}
We discuss the threats to validity, including internal threats and external threats in the section.

\subsection{Internal Threats}

\subsubsection{False Negatives of Detecting On-device Models}

App developers employ various techniques to safeguard on-device models from theft, such as encryption and obfuscation. 
In our study, we discover that some developers may slice the models and compile them into binaries, making it more challenging for third parties to detect them by semantic pattern matching. 
This may cause our model detection method to miss some models, resulting in false negatives. 
The current limitations of dynamic code analysis tools for iOS apps also hinder a more in-depth investigation of this area.
In this paper, we chose to invite experienced developers to manually search for encrypted or obfuscated on-device models in apps to mitigate this issue.

As a future direction, we plan to conduct a comprehensive analysis of the iOS app's code, including calling the model API, loading the model API, dynamically downloading the model API, and detecting the DL model from the code pattern level, to address the limitations of our current study.

\subsubsection{Bias for Model Functionality and Use Scenarios Inference}

We infer possible functions of the model in Section~\ref{sec:modelFunction} by leveraging semantic information of the app and the model. However, obtaining a comprehensive understanding of the usage of each model in an iOS app may require identifying the calling model at the code level and dynamically running the app to detect its usage. Current limitations of static and dynamic analysis tools for iOS apps hinder our ability to analyze models from the code level effectively as we do for Android apps. 
To mitigate possible inaccuracies in our functional inferences, we dynamically run the iOS app to verify our inferences. However, we cannot fully guarantee the accuracy of all inferences, resulting in potential bias in our current approach. 
In future work, we plan to explore alternative methods for analyzing the code of on-device models in iOS apps. This aims to reduce potential bias and improve the accuracy of our functional inferences.

\subsubsection{Impact of Remote Attacks on On-device Models in iOS Apps}

One potential threat to the validity of our study is the assumption that attackers have access to the targeted iOS devices and can perform the attacks locally. This may not be a realistic scenario in some cases, especially if the attack is conducted remotely.
Our study did not investigate the potential impact of remote attacks on on-device models in iOS apps, which is a limitation. Remote attacks may have different characteristics and may require different techniques than local attacks, and their effectiveness may depend on the network environment and other factors. Therefore, the results of our study may not fully reflect the security risks faced by on-device models in real-world scenarios.
However, we note that our proposed attack can still be effective in some scenarios where users need to capture inputs from the real-world, and the attacker can place adversarial inputs beforehand to cause misclassification. We also acknowledge that further research is needed to explore the impact of remote attacks on on-device models and to develop effective defenses against such attacks.

\subsubsection{Limitations on the scope of the attack method in RQ3}

Our attack method can only be applied to black-box models that have white-box counterparts, which represent 17.63\% of the 1,883 models discovered in iOS apps. We have only identified image-based models in real-world iOS apps, and lack experiments with models from other domains such as natural language processing and speech recognition. While our primary objective is to assess the robustness of current on-device models against adversarial attacks, our focus on image-based models limits the generalizability of our findings. 
It's worth noting that our proposed attack method can be easily migrated to other types of models, such as NLP and speech recognition. 
To address these limitations, we plan to design and analyze the effectiveness of our approach on more types of models algorithmically in future work. To expand the range of black-box models that we can attack, we plan to try to migrate the adversarial example to attack black-box models with similar structures and parameters.

\subsection{External Threats}
\subsubsection{Dynamic Analysis of On-device Models}
In this study, we primarily use static analysis to examine on-device models and code in Android and iOS apps. Although useful, static analysis cannot capture the state of on-device models after the app is launched. Dynamic analysis, on the other hand, can provide more in-depth insights into the current state of on-device models, including memory usage, call time analysis, and dynamic data security. Additionally, it can help collect more on-device models for obfuscated or encrypted models by analyzing the code of the app loading the model and dumping the model from the memory where the app is running. However, current dynamic analysis tools for iOS apps have limitations that prevent effective analysis of on-device models. As a result, we invite three experienced participants to manually analyze the running app to identify the app's functions and use scenarios that might call the on-device model. However, this approach proved less effective than direct dynamic analysis.

Therefore, we think that the development of better dynamic analysis tools for iOS apps is crucial for future research. These tools will enable more effective analysis of on-device models and facilitate further investigations on dynamic security risks, model attacks, and protection.

\subsubsection{Impact of Model Properties on Model Selection}

Model accuracy, interpretability, and explainability are important considerations when selecting on-device models. Accuracy is commonly used to evaluate model performance, while interpretability and explainability are crucial for building trust in the model and understanding how it makes predictions. However, balancing these factors can be challenging, as accuracy may come at the expense of interpretability and explainability, and vice versa. Moreover, the impact of these factors may vary across different application domains. While interpretability and explainability may be more critical in domains like healthcare and finance, accuracy may be the primary consideration in image or speech recognition. Despite their significance, the impact of these factors on developer model selection is often overlooked. 
The scope of this paper focuses on the impact of the differences between iOS and Android platforms on on-device model selection, so we have not delved into the impact of model properties on model selection in this paper. Still, it is a crucial area for future investigation.

\subsubsection{Impact of Platform Hardware and Software Configurations on Model Selection}

The differences in hardware and software configurations across platforms can influence the performance and compatibility of the on-device models, which may affect the model selection process. While this paper briefly touches upon the impact of hardware and software differences on model selection in Section~\ref{sec:suggestions} of RQ2, no detailed examination of how specific hardware and software factors affect the selection process is conducted. Thus, in future work, we aim to investigate the effects of hardware and software-specific factors on the model selection process.

\subsubsection{Impact of User Behavior and Preference on Model Selection}

Different apps have varying target users, the preferred model for one user may not necessarily be suitable for another. 
User preference may prioritize model accuracy or interpretability over computational resources or inference time. Moreover, users may have different concerns about data privacy and ethics, which can affect the selection and usage of specific models. 
Our study focuses on the perspective of app developers and DL framework providers when selecting on-device models and does not consider user behavior and preference. 
Future research will investigate the impact of user behavior and preference on model selection and usage to provide a more comprehensive understanding of on-device model adoption.

\section{Related Work}
Our work empirically studies the characteristics and potential security issues of DL models on iOS apps.
In this section, we discuss related works about deep learning techniques on mobile and adversarial attacks on deep learning models.

\subsection{Deep Learning Techniques on Mobile}
With the rapid development of computing hardware, more and more deep learning frameworks support the deployment of on-device models on mobile.
Deep learning technologies currently applied on smartphones can be divided into two categories: on-device model and on-cloud service.
Xu et al.~\cite{xu2019first} first explore the characteristics of deep learning apps on smartphones.
However, due to the difficulties of collecting iOS apps, they only study apps on Android, and some technologies like CNNDroid~\cite{latifi2016cnndroid}, Parrots~\cite{sensetime} and PyTorch~\cite{pytorch-mobile} are outdated.
Sim et al. investigate the effectiveness of on-device models on automatic speech recognition~\cite{sim2019investigation}.
Ravi et al.~\cite{ravi2019efficient} propose an approach to improve the effectiveness of on-device models on smartphones.
Their efforts exclusively address certain aspects of on-device models, and none of them include iOS apps.
In industry, more and more deep learning frameworks support on-device models on both Android and iOS, such as TF Lite, TensorFlow, Caffe2, Mace~\cite{mace}, and so on.
The increasing prevalence of the on-device model calls for a comprehensive study of it, particularly the on-device model on the iOS platform.
Some companies also provide the service of on-cloud deep learning models, like Google~\cite{googlecloud}, Amazon~\cite{amazoncloud}, and Microsoft Azure~\cite{Microsoftcloud}.
Prior studies focus on analyzing the machine learning as a service (MLaas) system~\cite{fredrikson2015model, tramer2016stealing, yao2017complexity, shokri2017membership, hu2019code}.
However, on-cloud models have disadvantages compared to on-device models, such as privacy concerns and unreliable accessibility. 
Consequently, some DL tasks are better suited for execution on local devices, such as face detection and photo beauty. 
Our research also demonstrates that some current apps use both on-device and on-cloud deep learning models for different tasks in one app.

\subsection{Adversarial Attacks on Deep Learning Models}
Adversarial attacks, which only add subtle perturbations to input images, can lead a deep learning model to output incorrect prediction results with high confidence level~\cite{zhang2019adversarial}.
Adversarial attacks have achieved remarkable progress, especially in the computer vision field.
Current adversarial attacks can be divided into two categories: white-box attacks and black-box attacks or targeted attacks and non-target attacks.
Huang et al.~\cite{huang2021robustness} first explore the robustness of on-device deep learning models on Android apps.
They propose a straightforward yet effective approach, which requires locating the pre-trained model of the on-device model, attacking the pre-trained model to generate adversarial examples, and then using the adversarial examples to attack the on-device model.
However, this work requires locating the pre-trained model and only investigating on-device models on iOS apps.
Since deep neural networks are vulnerable to adversarial attacks, researchers have proposed several defense methods, including adversarial training, transformation, distillation, and gradient regularisation~\cite{he2019towards, zhou2023modelobfuscator}.
Although several studies have proved the potential security issues of on-device models on Android apps, systemic study on the security of on-device models on iOS apps is scant.
Our work systematically investigates the security threat of the on-device model in the face of adversarial attacks in iOS Apps for the first time. It demonstrates that the On-device model of the Core ML framework in iOS Apps still poses significant security risks.

In prior research, some efforts~\cite{liu2016delving, fang2020local, tramer2017ensemble, huang2023training} have used an ensemble of multiple models to generate attacks that can be transferred over multiple models, and there is some overlap with our attack in RQ3 which uses white-box models as bridges. 
However, these approaches tend to be too theoretical and do not specifically target the on-device models in iOS apps for attack design, resulting in relatively low overall attack success rates. 
Our work is more practical and effective in real-world scenarios, as we generate adversarial examples against the target model based on its trainable form, which is more efficient than using an ensemble of multiple models. Additionally, we specifically design our attack for the unique characteristic of on-device models, taking into account the fact that backpropagation is disabled under this setting. This consideration enhances the practicality of our approach for on-device models in iOS apps.

\section{Conclusion}
This paper presents the first systematic empirical study on how real-world iOS apps exploit DL techniques.
We first explore the current characteristics of on-device models on iOS.
We find that some apps adopt different DL techniques across Android and iOS.
So, we further investigate why developers use different models between Android and iOS.
We propose a new approach to attack gray-box on-device models of Core ML on iOS, and the results illustrate the effectiveness of our approach.
Our future work will focus on how to attack gray-box on-device models without white-box counterparts and how to dynamically analyze on-device models when apps are running.

\bibliographystyle{ACM-Reference-Format}
\bibliography{reference}

\clearpage
\appendix

\end{document}